\documentclass[12pt,preprint]{aastex}

\usepackage{psfig}

\begin{document}
\title{The ROSAT All-Sky Survey: a Catalog of Clusters of Galaxies\\
in a Region of 1 Ster around the South Galactic Pole}
\author{R. Cruddace\altaffilmark{1,2},
W. Voges\altaffilmark{3}, H. B\"{o}hringer\altaffilmark{3},
C. A. Collins\altaffilmark{4}, A. K. Romer\altaffilmark{5},
H. MacGillivray\altaffilmark{6}, D. Yentis\altaffilmark{1}, 
P. Schuecker\altaffilmark{3}, H. Ebeling\altaffilmark{7},
and S. De Grandi\altaffilmark{8}}
\altaffiltext{1}{E. O. Hulburt Center for Space Research, Naval 
Research Laboratory, Washington DC 20375}
\altaffiltext{2}{Visiting Scientist, Max-Planck-Institut f\"{u}r 
Extraterrestrische Physik, 85740 Garching, Germany}
\altaffiltext{3}{Max-Planck-Institut f\"{u}r Extraterrestrische Physik, 85740 
Garching, Germany}
\altaffiltext{4}{Astrophysics Research Institute, Liverpool John
Moores University,Liverpool L3 5AF, England}
\altaffiltext{5}{Carnegie Mellon University, 5000 Forbes Avenue,Pittsburgh, 
Pennsylvania 15213}
\altaffiltext{6}{Institute of Astronomy, University of Edinburgh, 
Blackford Hill, Edinburgh EH9 3HJ, Scotland}
\altaffiltext{7}{Institute for Astronomy, University of Hawaii, 2680 
Woodlawn Drive, Honolulu, Hawaii 96822} 
\altaffiltext{8}{Osservatorio Astronomico di Brera, via Bianchi 46, 
22055 Merate (LC), Italy}

\begin{abstract}
A field of 1.013 ster in the ROSAT all-sky survey (RASS), centered on
the south galactic pole (SGP), has been searched in a systematic, 
objective manner for clusters of galaxies. The procedure relied on a 
correlation of the X-ray positions and properties of ROSAT sources in 
the field with the distribution of galaxies in the COSMOS digitised 
data base, which was obtained from scanning the plates of the UK 
Schmidt IIIa-J optical survey of the southern sky. The study used the 
second ROSAT survey data base (RASS-2) and included several optical 
observing campaigns to measure cluster redshifts. The search, which 
is a precursor to the larger REFLEX survey encompassing the whole 
southern sky, reached the detection limits of both the RASS and the 
COSMOS data, and yielded a catalog of 186 clusters in which the 
lowest flux is  
$1.5\times 10^{-12}\,{\rm erg}\,{\rm cm}^{-2}\,{\rm s}^{-1}$ in the 
$0.1-2.4\,{\rm keV}$ band. 
Of these 157 have measured redshifts. Using a flux limit of 
$3.0\times 10^{-12}\,{\rm erg}\,{\rm cm}^{-2}\,{\rm s}^{-1}$
a complete subset 
of 112 clusters was obtained, of which 110 have measured redshifts.
The spatial distribution of the X-ray clusters out to a redshift of 
$0.15$ shows an extension of the Local Supercluster to the Pisces-Cetus
supercluster $({\rm z}<\sim 0.07)$, and an orthogonal structure at higher 
redshift $(0.07<{\rm z}<0.15)$. This result is consistent with large-scale 
structure suggested by optical surveys.
\end{abstract}

\section*{INTRODUCTION}
It has been recognized for many years that observations of clusters of galaxies
can be applied effectively to test theories of the evolution of the 
universe. As the largest gravitationally bound objects emerging from the growth
and subsequent collapse of primordial density fluctuations, their spatial
distribution at different epochs is a relic containing evidence which can
test cosmological models of structure formation. X-ray astronomy has 
enhanced the results of such studies greatly, for a number of reasons. First,
clusters are X-ray sources of relatively high luminosity, with a luminosity
function sufficiently well defined to allow, in principle at least, cosmic 
structure to be studied out to epochs of ${\rm z}\sim 1$ and beyond. Further,
ambiguities in the identification of clusters, caused by chance line-of-sight
superposition of objects, are fewer in X-ray than in optical sky surveys. This
is a consequence of the sharply peaked X-ray brightness distribution in most
clusters, and of the relatively low surface density of bright X-ray sources.
Second, the X-ray
properties of clusters are defined by events reaching back to the earliest
epoch of cluster formation, and therefore may be used to test cosmological 
models. The X-ray luminosity and temperature of the intracluster gas depend on
the cluster mass, so that the shape and amplitude of the luminosity and 
temperature functions 
of the cluster population are dependent on the spectrum of the early density
perturbations. Further, the morphology of clusters in the whole population may
be dependent on the characteristics of the expansion, and is a diagnostic of 
cluster merging processes. Finally, the abundance of heavy elements in the 
intracluster gas is a measure of galaxy evolution and ram-pressure stripping, 
and of the proportion of primordial gas in the cluster.

Therefore much effort has been given to the production of statistically complete
samples of X-ray emitting clusters. The first were produced using $\it Uhuru$
(Schwartz 1978) and $\it Ariel\ V$ (McHardy 1978) observations, and the HEAO-1
all-sky surveys (Piccinotti et al. 1982; Kowalski et al. 1984), and contained 
up to 76 clusters. The flux limits of these surveys 
($1-8\times 10^{-11}\,{\rm erg}\,{\rm cm}^{-2}\,{\rm s}^{-1}$, expressed here 
in the ROSAT band $0.1-2.4\,{\rm keV}$), and their angular resolution were 
limited by the 
use of proportional counters, and improvements were possible only after the 
first missions using imaging X-ray instruments. Edge et al. (1990) reported the 
use of EINSTEIN and EXOSAT observations to refine the results of the 
earlier surveys, producing a sample of 55 clusters free of, for example, the 
effects of source confusion. In the EINSTEIN Extended Medium Sensitivity 
Survey (EMSS) it was possible, by examining the fields of EINSTEIN pointings
covering a total of 740 ${\rm deg}^{2}$, to create a sample with a flux limit 
of $1.3\times 10^{-13}\,{\rm erg}\,{\rm cm}^{-2}\,{\rm s}^{-1}\,
(0.3-3.5\,{\rm keV})$ and containing 
clusters with redshifts out to ${\rm z}=0.58$ (Henry et al. 1992).

The ROSAT X-ray telescope, equipped with an imaging proportional counter in
the focal plane, allowed for the first time complete detailed surveys of large 
areas of sky (Tr\"{u}mper 1983). The analysis of the ROSAT All-Sky Survey 
(RASS) proceeded in two phases RASS-1 and RASS-2, described below in section 
3.1, in which the second phase incorporated a number of improvements 
suggested by RASS-1 (Voges et al. 1999). The first X-ray flux-limited samples 
of clusters based on analysis of a large area of sky in the ROSAT survey were
obtained by Romer et al.(1994), Romer (1995), Ebeling et al.(1996) and Ebeling 
et al. (1998). The SGP sample described in Romer et al.(1994), based on 
analysis of
the RASS-1 data base, contained 161 cluster candidates with a limiting flux of 
$10^{-12}\,{\rm erg}\,{\rm cm}^{-2}\,{\rm s}^{-1}\,(0.1-2.4\,{\rm keV})$. 
Redshifts were obtained for 128 clusters in this sample,
and this subset was used in a study of the cluster spatial correlation function
(Romer et al. 1994). The BCS survey by Ebeling et al. (1998) addressed the 
whole extragalactic sky (${\rm b}\geq 20^{o}$) in the northern hemisphere, and 
reanalysed 
the fields in which clusters had been detected in the RASS-1 analysis. The 
survey yielded a sample of 201 clusters at ${\rm z}\le 0.3$, having 
a limiting flux of $4.4\times 10^{-12}\,{\rm erg}\,{\rm cm}^{-2}\,{\rm s}^{-1}
\,(0.1-2.4\,{\rm keV})$, which is statistically complete at the 90 percent
level. In a recent publication the BCS sample has been extended to an X-ray 
flux limit of $2.8\times 10^{-12}\,{\rm erg}\,{\rm cm}^{-2}\,{\rm s}^{-1}$, 
at which the completeness was 75 percent (Ebeling et al. 2000). A recent survey
of clusters in the 
northern sky (NORAS), based on X-ray sources classified as extended in the RASS
analysis, is described by B\"{o}hringer et al. (2000). Deeper surveys have been 
possible in areas around the ecliptic poles (Henry et al. 2001, Gioia et al.
2001, Mullis et al. 2001 and Voges et al. 2001), where the ROSAT survey
achieved unusually long exposure times, and in areas where deep pointings have 
been made (Romer et al. 2000, Jones et al. 1998, Rosati et al. 1998, 
Vikhlinen et al. 1998 and Collins et al. 1997). The deepest survey was made by 
Rosati et al. (1998), in which the sample was complete down to a flux of 
$4\times 10^{-14}\,{\rm erg}\,{\rm cm}^{-2}\,{\rm s}^{-1}$ and included clusters
out a redshift of ${\rm z}\sim 0.8$.
                                           
In what follows we concentrate on complete large-area cluster surveys in the 
southern hemisphere. This has its roots in an agreement between the Royal 
Observatory in Edinburgh (ROE), the Naval Research Laboratory (NRL) and the 
Max-Planck-Institut f\"{u}r Extraterrestrische Physik (MPE) in Garching, made 
early in the ROSAT program,
to make the COSMOS digitised UK Schmidt IIIa-J survey of the southern sky 
(MacGillivray \& Stobie 1985, Yentis et al. 1992) available for identification 
of X-ray sources. This data base, containing $3.5\times 10^{8}$ objects with a 
limiting magnitude of ${\rm b}_{j}\sim 22$, was installed at MPE for use in 
X-ray source identification. In cluster 
studies it was used to derive unbiased lists of candidates through systematic 
correlations of X-ray source positions with concentrations of galaxies. A
comprehensive search for clusters is being undertaken at MPE in the REFLEX 
southern hemisphere project (B\"{o}hringer et al. 2001). The first 
results (De Grandi et al. 1999) described a sample of X-ray bright clusters, 
obtained by searching a region of area 
$8235\,{\rm deg}^{2}$ ($2.5\,{\rm ster}$). The study reanalysed clusters 
identified
in the RASS-1 analysis and arrived at a sample of 130 clusters having a 
completeness of at least 90 percent and a limiting flux of 
$5-6.6\times 10^{-12}\,{\rm erg}\,{\rm cm}^{-2}\,{\rm s}^{-1}$ in the 
$0.1-2.4\,{\rm keV}$ band. 
 The complete REFLEX study area covers $4.24\,{\rm ster}$ in the southern sky,
and the techniques used in constructing the cluster sample have been described
by B\"{o}hringer et al. (2001). At the current stage of the study a sample of
452 clusters has been obtained, which is at least 90 percent complete, using a 
flux limit of $3\times 10^{-12}\,{\rm erg}\,{\rm cm}^{-2}\,{\rm s}^{-1}$.
A power-law fit to the log N-log S distribution yielded an exponent of
${\rm \alpha }=-1.39\pm 0.07$.
Collins et al. (2000) have analysed the spatial correlation function of
clusters using a REFLEX sample of 452 clusters covering an area 4 times larger
than the SGP, and derived a value of $18.8\pm 0.9\,{\rm h}^{-1}$ Mpc for the 
correlation length, significantly greater than the value of 
$12.9\pm 2.2\,{\rm h}^{-1}$ Mpc obtained by Romer et al. (1994). However for 
the 109
REFLEX clusters in the SGP area Collins et al. (2000) find a correlation length
of $12.9\pm 1.9\,{\rm h}^{-1}$ Mpc, consistent with the result of Romer et al. 
(1994).
The difference between the complete REFLEX and the SGP results is most likely
due to sample fluctuations caused by cosmic variance (see Collins et al. 2000).
A significant conclusion of the REFLEX study is that low-density cold dark 
matter (CDM) models (${\rm \Omega }\sim 0.3$) provide a better fit to the 
observed 
correlation function, and that high-density CDM models (${\rm \Omega }\sim 1$) 
yield a poorer fit because they provide insufficient power at large scales.
Schuecker et al. (2001) have analysed the power spectrum of the REFLEX sample,
and the results are consistent with the results of Collins et al. (2000).

A prominent
aspect of the SGP and REFLEX survey projects has been optical observing 
programs to obtain cluster redshifts, which we describe briefly in section 2. 
However, it should be borne in mind that while these two projects have a common 
root and have followed similar procedures for cluster selection, X-ray analysis 
and redshift determination, they do possess some differences in approach. 

Following this introduction section 2 provides a summary of how the SGP survey
was performed. Section 3 describes the various procedures by which the cluster
sample was extracted from the RASS-2 data base and section 4 summarises how the
X-ray energy flux and luminosity were calculated. Section 5 presents the SGP 
cluster catalog in the form of a table and describes the archive of SGP
cluster images in the form of overlays of the X-ray and optical data, which is
accessible at MPE over the Internet. Section 6 examines the completeness of the 
cluster sample and section 7 discusses the large-scale structure traced by the 
clusters out to a redshift of ${\rm z}\sim 0.15$. Finally a summary of the 
results and conclusions of the paper are presented in section 8. 
Throughout the article we have assumed that the Hubble constant 
${\rm H}_{o}=50\,{\rm km}\,{\rm s}^{-1}\,{\rm Mpc}^{-1}$, the deceleration 
parameter ${\rm q}_{o}=0.5$ and the cosmological constant ${\rm \Lambda }=0$.
\section*{2. SUMMARY OF THE SOUTH GALACTIC POLE SURVEY}
This survey selected a region at high galactic latitude in the southern
sky centered on the South Galactic Pole. Rectangular in shape, it extends in 
right ascension from $22^{\rm h}$ to $3^{\rm h}\ 20^{\rm m}$ and in 
declination from 
$-50^{\rm o}$ to $+2.5^{\rm o}$ (see Figure \ref{f18}), giving it an area of
1.013 steradian. The search for clusters in this region invoked no flux 
threshold, but intead worked to the limit of the RASS data to maximise the
completeness of the sample of cluster candidates. A flux limit is established
subsequently to define a complete subsample.

The first step in the search for clusters was to select all X-ray sources found
in the region by the RASS-2 analysis (Voges et al. 1999). In this analysis the
threshold values for the source detection algorithmns were lowered 
(section 3.1), with the aim of achieving as deep a survey as 
the RASS would allow. Therefore cluster searches using this data base have more
sensitivity than has been available to previous southern hemisphere searches 
using the RASS-1 data base (Romer 1995, De Grandi et al. 1999). 

The RASS-2 analysis was not equipped for detailed examination of the X-ray
properties of extended sources such as clusters. To this end a number of
techniques have been developed, including Voronoi Tesselation and Percolation 
(VTP, Ebeling \& Wiedemann 1993), the Steepness Ratio Technique 
(SRT, DeGrandi et al. 1997) and the Growth Curve Analysis (GCA, 
B\"{o}hringer et al. 2000). The goal of these techniques is to obtain
reliable fluxes for both extended and point sources, and to extract from the raw
data as much information as possible about the X-ray source characteristics. In
this study the GCA technique has been used to analyse the list of 
candidates obtained from the RASS-2 data base.

Cluster candidates were selected from this sample using the COSMOS digitised
IIIa-J survey and a search procedure called CSEARCH. This procedure, which is
not designed to select clusters in an unambiguous manner, compares the
galaxy density around the RASS-2 position with the background galaxy density on 
the UK Schmidt plate, and derives a probability that the result was a random 
coincidence. A sample of cluster candidates is obtained by setting a threshold
to this probability, and this threshold is set low enough to ensure that few
genuine clusters are lost and a high completeness is attained in the final 
sample. In our study a RASS-2 list of 11981 sources is reduced by CSEARCH to 
3236 candidates (section 3.3). At this point much remains to be done to 
remove contaminating sources, which we estimate from internal statistical 
tests to comprise about half the candidates, in an objective process. At high 
redshifts (${\rm z}>\sim 0.3$) the method reaches a limit of the COSMOS data,
when many cluster galaxies become so faint as to be classified by COSMOS only
as 'faint objects', which are not counteed by CSEARCH. Therefore some clusters
at high redshift may be missed, which may be one source of incompleteness 
near the flux limit of the survey.

The analysis of this sample of cluster candidates began by setting a limit of
$0.08\,{\rm ct}\,{\rm s}^{-1}$ to the X-ray count rate in the hard band
($0.5-2.0\,{\rm keV}$, section 3.1, 3.4 ), which reduced the number of 
candidates from 3236 to 477. This was followed by the process of removing 
contaminating sources from this sub-sample, which proceeded through a number
of stages. The first was the removal of COSMOS artifacts, bright sources which 
to the eye were clearly not clusters, and redundant detections by RASS-2 of the 
same source, particularly in extended sources such as clusters. This was done 
by examination of COSMOS optical finding charts and of overlays of the optical 
and X-ray
images. The second was to correlate RASS-2 positions with the SIMBAD
catalog, in order to eliminate bright stars found in the RASS-2 error circle,
and then to use two X-ray properties of the source, hardness ratio and
extent, as selection criteria. The hardness ratio was used to exclude those 
stars and active galactic nuclei (AGN) whose spectra are sufficiently soft to
exclude the possibility that the source is a cluster. The X-ray extent was used
to classify candidates in three groups, namely
clusters, AGN candidates and a remainder of uncertain nature. The third and
final stage, identifying the clusters and known AGN among the remaining
sample, proceeded along two parallel
paths. The first comprised correlations of the candidate list with a variety of
catalogs, in particular the NASA Extragalactic Data base (NED), the 
SIMBAD catalog, and the Veron catalog of AGN (Veron-Cetty \&  Veron 1998). The 
second was an ongoing examination of the results of the supporting optical 
observations, described below, which identified AGN in the sample.

We summarise here briefly the sources of incompleteness in the final catalog 
obtained, which reduce the cluster count at low fluxes:\\
\hspace*{1cm}i) The CSEARCH analysis has a predetermined level of incompleteness
in correlating\\
\hspace*{1.5cm}X-ray sources with galaxy concentrations (section 3.3).\\
\hspace*{1cm}ii) The sharp count-rate limit ($0.08\,{\rm ct}\,{\rm s}^{-1}$) 
rejects some sources at low energy\\
\hspace*{1.5cm}flux, due to the variation of interstellar column density in the
field (section 3.4).\\
\hspace*{1cm}iii) Some distant clusters elude the tests applied in searching the
list of candidates\\
\hspace*{1.5cm}yielded by CSEARCH and the X-ray count-rate cut (Section 3.5).\\

Finally, a vital complementary part of the SGP survey and
the REFLEX project has 
been a program of optical observations to obtain the redshifts of those clusters
detected in the ROSAT survey, for which no redshifts were available in the
literature. The SGP survey, which commenced not long after the launch of ROSAT 
in 1990, was supported by three campaigns at the 4m Anglo-Australian Telescope
(AAT) and three at the 1.9m telescope of the South African Astronomical 
Observatory. The results of these campaigns have been reported by Romer (1995). 
A larger program, which started in 1992 and was completed in 1995, was 
installed as a Key Project at the European Southern Observatory (ESO). 
Its purpose was to obtain redshifts for clusters in the REFLEX sample. An 
ongoing continuation of these observations is being pursued to support the
extension of the REFLEX survey to lower X-ray flux limits. 
\section*{3. SELECTION OF THE CLUSTER SAMPLE.}
\subsection*{3.1 The X-ray Data Base.}
The ROSAT observatory was launched on 1990 June 1, and saw first light on 1990 
June 16 (Tr\"{u}mper et al. 1991). After a two-month verification of the 
instrument operation it commenced its six-month long survey of the whole sky on
30 July 1990. During the RASS the observatory rotated about an axis through the
sun, scanning on a given day a $2{^{\rm o}}$ wide strip passing through the 
ecliptic 
poles. The survey ended on 1991 February 18. The first analysis of the survey 
(RASS-1), which took place in 1991-1993, used data collected into the same 
$2^{\rm o}$ wide strips, and yielded a list of some 50,000 sources containing
positions and such characteristics as count-rate, hardness ratio, angular
extent and source detection likelihood. RASS-1 was used to test the algorithms
implemented in the analysis and to start scientific studies, and as a 
consequence a series of improvements was initiated (Voges et al. 1999):\\ 
\hspace*{1cm}i) Sorting the data into 1378 sky fields of size
$6^{\rm o}.4\times 6^{\rm o}.4$, overlapping at least $0.23^{\rm o}$.\\
\hspace*{1.5cm}This improved the efficiency of source detection.\\
\hspace*{1cm}ii) For operational reasons the RASS-1 data base contained a few
regions of low \\ 
\hspace*{1.5cm}exposure, which were filled in later by supplementary 
observations. This \\
\hspace*{1.5cm}later data was included in the RASS-2 data base.\\
\hspace*{1cm}iii) Searching for sources in the soft $(0.1-0.4\,{\rm keV})$, 
hard $(0.5-2.0\,{\rm keV})$ and broad\\
\hspace*{1.5cm}$(0.1-2.4\,{\rm keV})$ energy bands independently.\\
\hspace*{1cm}iv) Refining the pointing attitude solution and improving source 
positions.\\
\hspace*{1cm}v) Improvement of the spline-fit to the background, yielding 
more accurate source\\
\hspace*{1.5cm} count-rates.\\ 
\hspace*{1cm}vi) The list of candidates selected by the sliding-window source 
detection algorithmns\\
\hspace*{1.5cm}was increased by lowering the detection threshold, and in the
subsequent maximum\\
\hspace*{1.5cm}likelihood analysis the likelihood threshold for accepting
sources was lowered\\
\hspace*{1.5cm}from 10 (RASS-1 analysis) to 7.\\ 

Reanalysis of the survey data using the new software (RASS-2) was performed in
1994-1995, yielding a list of 145060 sources. The SGP field contained 11981
sources above a likelihood threshold of 7. The characteristics of the new
software and results from the RASS-2 analysis have been described by 
Voges et al. (1999). The results described here relied solely on the RASS-2 
data base (RASS bright source catalog in Voges et al. 1999 and RASS faint source
catalog in Voges et al. 2000).

\begin{figure*}
\centerline{\psfig{figure=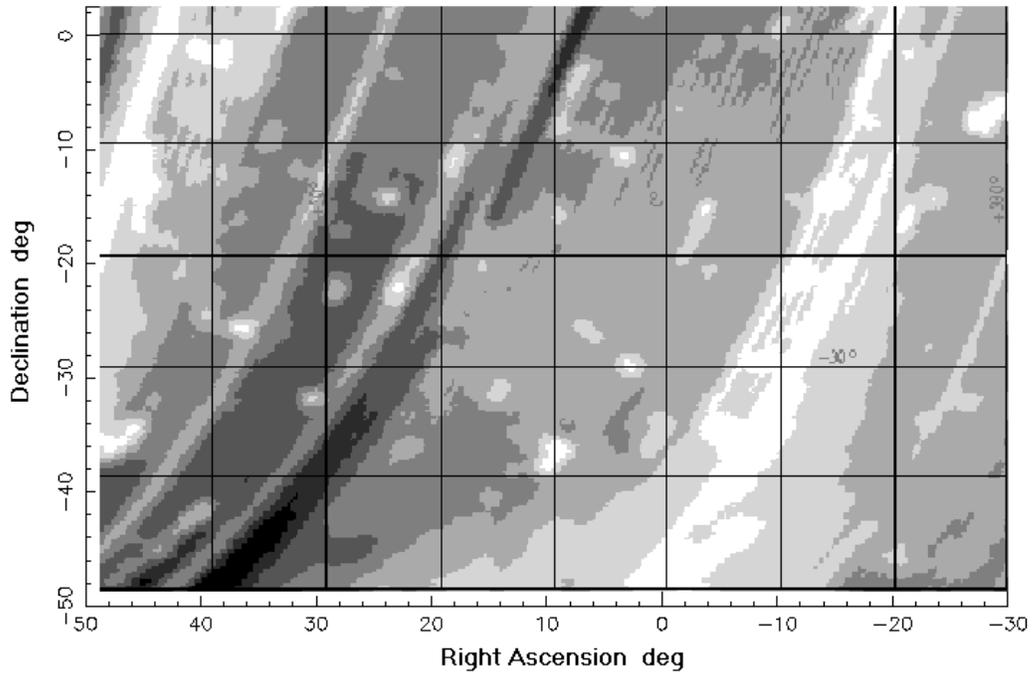,width=6.0in}}
\caption{A map of the ROSAT survey exposure time distribution in the South
Galactic Pole survey region. The increase toward the SE corner is a consequence
of the procedure by which ROSAT scanned the sky (see text), and the striations 
are the result of shutting down the detector during passage through radiation
belts.}
\label{f1}
\end{figure*}

\begin{figure*}
\centerline{\psfig{figure=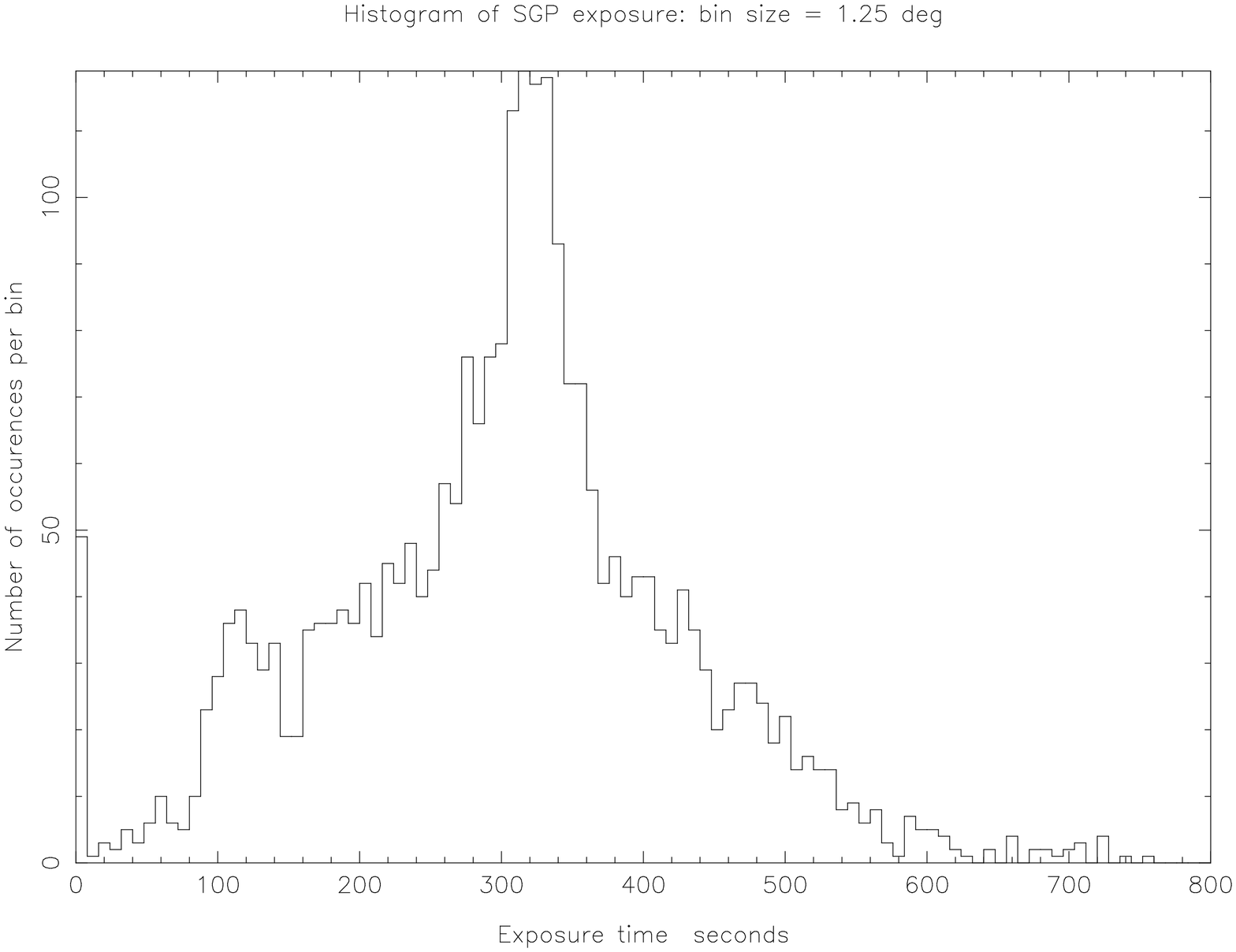,width=4.0in,angle=270}}
\caption{A histogram of the exposure times achieved in the SGP survey region, as
displayed in the exposure map in Figure \ref{f1}.}
\label{f2}
\end{figure*}

The $2^{\rm o}$ wide RASS scan strips overlap to an extent which increases with 
ecliptic latitude, and consequently the exposure time grows and reaches a 
maximum at the poles. This effect is visible in the map of the RASS
exposure time in the SGP survey region (Figure \ref{f1}), in which there is a
systematic increase from the NW to the SE corner. Upon this change are
superposed striations caused by periodic protective shutdown of the 
focal-plane detector as it traversed radiation belts. Analysis of the exposure
map of the SGP survey region yields the histogram shown in
Figure \ref{f2}, in which the mean exposure time is 300 seconds. There is
an area of $75\,{\rm sq}\,{\rm deg}$ in the SGP field, in which the exposure 
time is less than $10\,{\rm s}$.

\subsection*{3.2 The COSMOS Data Base.}
The ROE COSMOS digital data base contains the results of scanning 894 plates 
of the UK Schmidt IIIa-J (blue) southern hemisphere survey. The SGP study area 
contains 161 fields, each of size $5^{\rm o}.35\times 5^{\rm o}.35$. COSMOS 
analyses and
classifies each object in a field, and records the ${\rm b}_{j}$ magnitude, 
position (R.A. and declination (2000)) and object shape parameters. Objects are
classified primarily as stars, galaxies or faint objects, the last being objects
whose low intensity on the plate defies a reliable determination of their class.
Heydon-Dumbleton, Collins, \&  MacGillivray (1989) compared the COSMOS galaxy 
number count-magnitude
distribution with those derived from several independent optical surveys, and
deduced that the galaxy sample was essentially complete ($\sim 95$ percent 
completeness level) to a magnitude limit ${\rm b}_{j}=20.0$. Beyond this limit 
the star-galaxy discrimination gradually becomes less reliable as ${\rm b}_{j}$
increases (MacGillivray \&  Stobie 1985, Heydon-Dumbleton et al. 1989), leading
to a significant incompleteness of the galaxy sample in the range 
${\rm b}_{j}>20$. In the COSMOS data base star/galaxy separation was performed 
down to a limit of ${\rm b}_{j}\sim 21.0$. An example of the results is
given in a figure, which shows the $\sim 20000$ galaxies identified by 
COSMOS in the SGP field UKJ 411 (this figure is in the published article, but
was too large for this archived version).

The galaxy magnitudes have been calibrated using photometric CCD observations of
sequences of galaxies, which were used also in the calibration of the 
Edinburgh/Durham galaxy catalog (Heydon-Dumbleton et al. 1989).
For any given plate COSMOS yields a 
galaxy intensity, which depends on such factors as exposure time, emulsion
characteristics and measures taken to keep the plate in a dry environment during
exposure. This implies that the galaxy magnitudes for any given plate must be
adjusted to match the surrounding plates. This is done by identifying galaxies
in the overlapping border zone of adjacent plates, and deriving a shift which 
matches the galaxy magnitudes. No unique solution for the whole southern sky is
possible, and instead a set of plate magnitude shifts is obtained by a fitting
procedure which globally minimises the inconsistencies between adjacent plates.
The resulting 'COSMOS galaxy magnitudes' are on an arbitrary machine scale,
which now must be adjusted to match the CCD calibrations, in essence providing
a global zero point solution to the matched plate data. This is done by
fitting a correlation between the CCD and COSMOS magnitudes of galaxies in
the CCD observation sequences. 

To further ensure there were no calibration problems as a result of the 
iosophotal thresholding of the COSMOS galaxies (isophotal thresholding results
in loss of light below the isophote level used and hence underestimation of 
the true brightness of a galaxy), we accepted for calibration purposes only 
galaxies in the COSMOS data which were at least three magnitudes above the 
plate limit (i.e. ${\rm b}_j<19.0$).
This resulted in 207 sequence galaxies being accepted for calibration. The
RMS scatter of the correlation between the COSMOS and the CCD magnitudes for
these galaxies was found to be $0.25$. As the intrinsic error in the CCD 
photometry is estimated to be $0.1-0.2$, we estimate that the resulting
error in the magnitudes derived for galaxies in the COSMOS IIIa-J southern 
hemisphere survey is $\sim 0.2$.   

Some independent corroboration of the COSMOS galaxy photometry is provided by
the study of Caretta, Maia, \&  Willmer (2000). They confirm after comparisons 
with their 
CCD data that the COSMOS galaxy magnitudes are reliable at the $\sim 0.2$ level.
Their results indicate that galaxy samples are $90$ percent complete at 
magnitudes down to ${\rm b}_{j}=19.5-20.0$, in broad agreement with the COSMOS
estimates. However this agreement should be treated with caution, as Caretta et 
al. (2000) studied a small area of sky having a lower galaxy count than is
available in the UK Schmidt southern hemisphere survey.
\subsection*{3.3 CSEARCH: the Selection of Cluster Candidates.}
The cluster search procedure (CSEARCH) counts galaxies within 5 circles of 
radius ${\rm R}$ = 1.5, 3, 5, 7.5 and 10 arcmin (circles 1 through 5), 
centered at the 
ROSAT source position, and compares the results with those of random sampling 
in the UKJ field. These radii correspond to the range of expected cluster core
diameters (e.g. at ${\rm z}\sim 0.3$ a radius of 300 kpc is equivalent to an 
angular scale of 1.1 arcmin and at ${\rm z}\sim 0.03$ to 11 arcmin). The 
random sampling was performed 1000 times for each radius, 
yielding for each field five histograms of the galaxy count. For each histogram
a table is created, giving the probability ${\rm p}$ that the galaxy count 
exceeds a value, ${\rm N}_{g}^{*}$:\\
\begin{equation}
{\rm p}=\frac{\Sigma^{\infty }_{N^{\ast}_{g}}{\rm n}({\rm N}_{g})}
{\Sigma^{^\infty }_{0}{\rm n}({\rm N}_{g})}
\end{equation}
where ${\rm n}$ is the number of cases where the galaxy count was 
${\rm N}_{g}$. Selection 
of cluster candidates is made after setting a threshold ${\rm p}={\rm p}_{t}$, 
which implies
that for each plate a threshold be set for ${\rm N}_{g}^{*}$. Individual plate 
sensitivity and statistics make this galaxy count threshold vary from plate to 
plate and yield a significant uncertainty. The following argument describes how
this is circumvented by analysing the ensemble of probability values ${\rm p}$ 
for the whole SGP RASS-2 source sample.

\begin{figure*}
\centerline{\psfig{figure=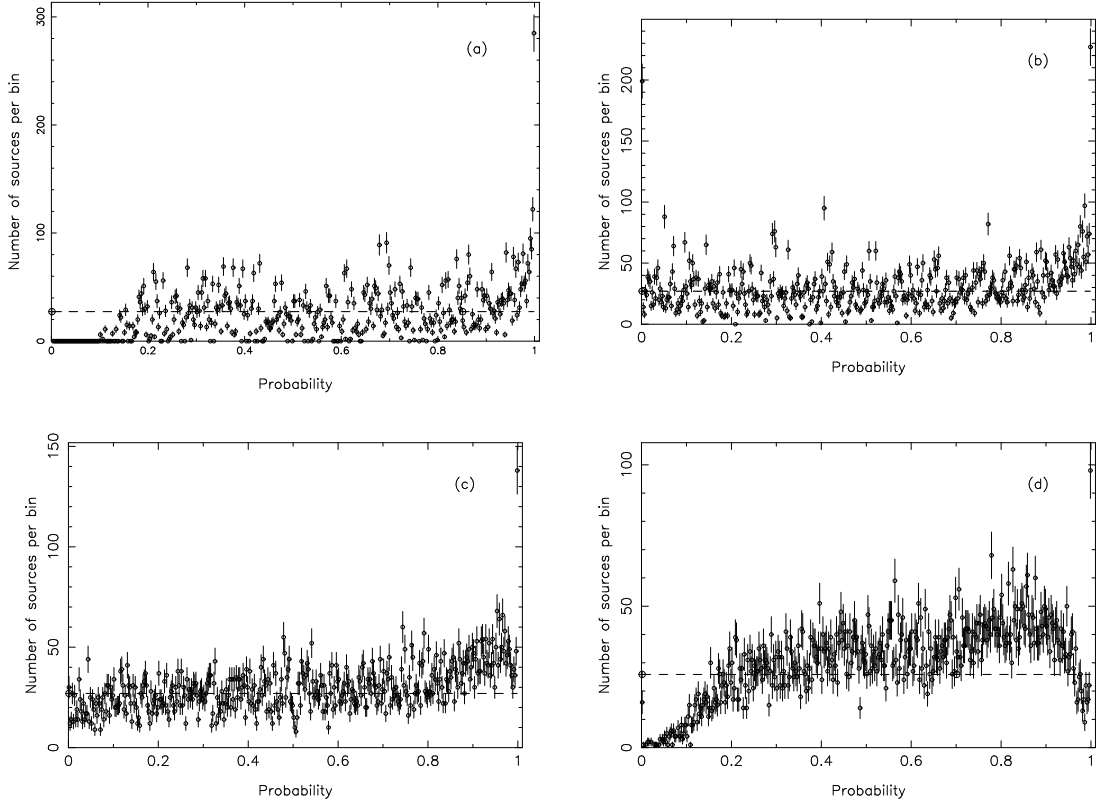,width=6.0in,angle=270}}
\caption{The CSEARCH analysis counts galaxies within a circle of specified
radius centered on the ROSAT source position, compares this with the results
of 1000 random samplings over each UK Schmidt field, and computes a probability
${\rm p}$ that the measured count is a statistical fluctuation of the 
backgound. We define a quantity ${\rm P}=1-{\rm p}$ which tends to 1 in dense 
galaxy concentrations. and divide the range $0<{\rm P}<1$ into 400 bins of 
equal width. The four plots show the distributions of source count per bin,
using all ROSAT sources in the SGP region, obtained using circles of radius 
(a) $1.5$, (b) $3$, (c) $5$ and (d) $10$ arcmin.}
\label{f4}
\end{figure*}

The ROSAT source is accepted if the probability associated with one of the 
5 galaxy counts (circles 1 through 5) at the ROSAT position falls below a 
selected threshold ${\rm p}_{t}$. For convenience, we have defined and used the 
quantity ${\rm P}=1-{\rm p}$, as this approaches a maximum value of 1 for 
cluster candidates. The range $0<{\rm P}<1$ is divided into bins of equal width
(typically 400 in number), and for any given search radius ${\rm R}$ we create 
a histogram of the source count (${\rm n}_{x}$) per bin for all the RASS-2 
sources in the SGP field. If there were no correlation 
between the ROSAT position and the galaxy count, as would be close to the truth
for an ensemble of stars and AGN, the expectation value for the number of 
sources, ${\rm n}_{x}$, per bin would be independent of ${\rm P}$ 
(B\"{o}hringer et al. 2001). Figure \ref{f4} shows the histograms of the
probability P obtained for the ROSAT sources detected in the SGP region. To 
save space we show results for ${\rm R}$ = 1.5, 3, 5 and 10 arcmin only. The 
major characteristic of these histograms is indeed a flat distribution with
statistical fluctuations, but in addition there is an enhancement at values of 
${\rm P}$ approaching $1$, which is the signature of clusters.

A comment on the data used to create these histograms is necessary. Following 
the first CSEARCH analysis of the SGP sources in the RASS-2 data base, 
examination of images overlaying the X-ray and the optical fields revealed 
multiple detections by RASS-2 of the more nearby clusters, due to their markedly
extended X-ray emission (e.g. A119). This biases the CSEARCH analysis, and 
therefore we removed 256 redundant multiple detections from the RASS-2 SGP 
source list, and then repeated the analysis. This number is small compared to
the total of 11981 sources, but is more significant in relation to the size of
the eventual CSEARCH sample of cluster candidates (3236).

The threshold ${\rm P}_{t}$ must be chosen carefully in terms of the desired 
compromise between formal completeness ${\rm C}_{o}$ and contamination 
${\rm C}_{t}$, for 
inevitably the selected sample contains ROSAT sources having chance 
associations with concentrations of galaxies, which build the flat 
distribution, and so the price of higher completeness is an increased 
contamination. The probability threshold ${\rm P}_{t}$ is related to 
${\rm C}_{o}$ and ${\rm C}_{t}$ by the following equations:\\ 

\begin{equation}
{\rm C}_{o}=\frac{\Sigma^{1}_{P_{t}}({\rm n}_{x}-{\rm n}_{b})}
{\Sigma^{1}_{0}({\rm n}_{x}-{\rm n}_{b})}
\end{equation}
\begin{equation}
{\rm C}_{t}=\frac{\Sigma^{1}_{P_{t}}{\rm n}_{b}}{\Sigma^{1}_{P{_t}}{\rm n}_{x}}
\end{equation}
where ${\rm n}_{b}$ is the number of sources per bin in a specified flat 
portion of the probability histogram.
       
Before the results of CSEARCH are discussed, two characteristics of the
histograms shown in Figure \ref{f4} should be clarified. First, the dip
in ${\rm P}$ for circle 1 in the range $0<{\rm P}<0.15$ is a quantisation 
effect, due to 
the galaxy count jumping from 0 to 1. The dip for circle 5 in the same region 
appears to be due to sampling limits, resulting from the circle diameter being 
a significant fraction of the field width. We have examined the number density
of clusters found in our analyses as a function of position on the plate, and
have found no sign of incompleteness near the COSMOS digitisation boundaries on
the plate. 

\begin{figure*}
\centerline{\psfig{figure=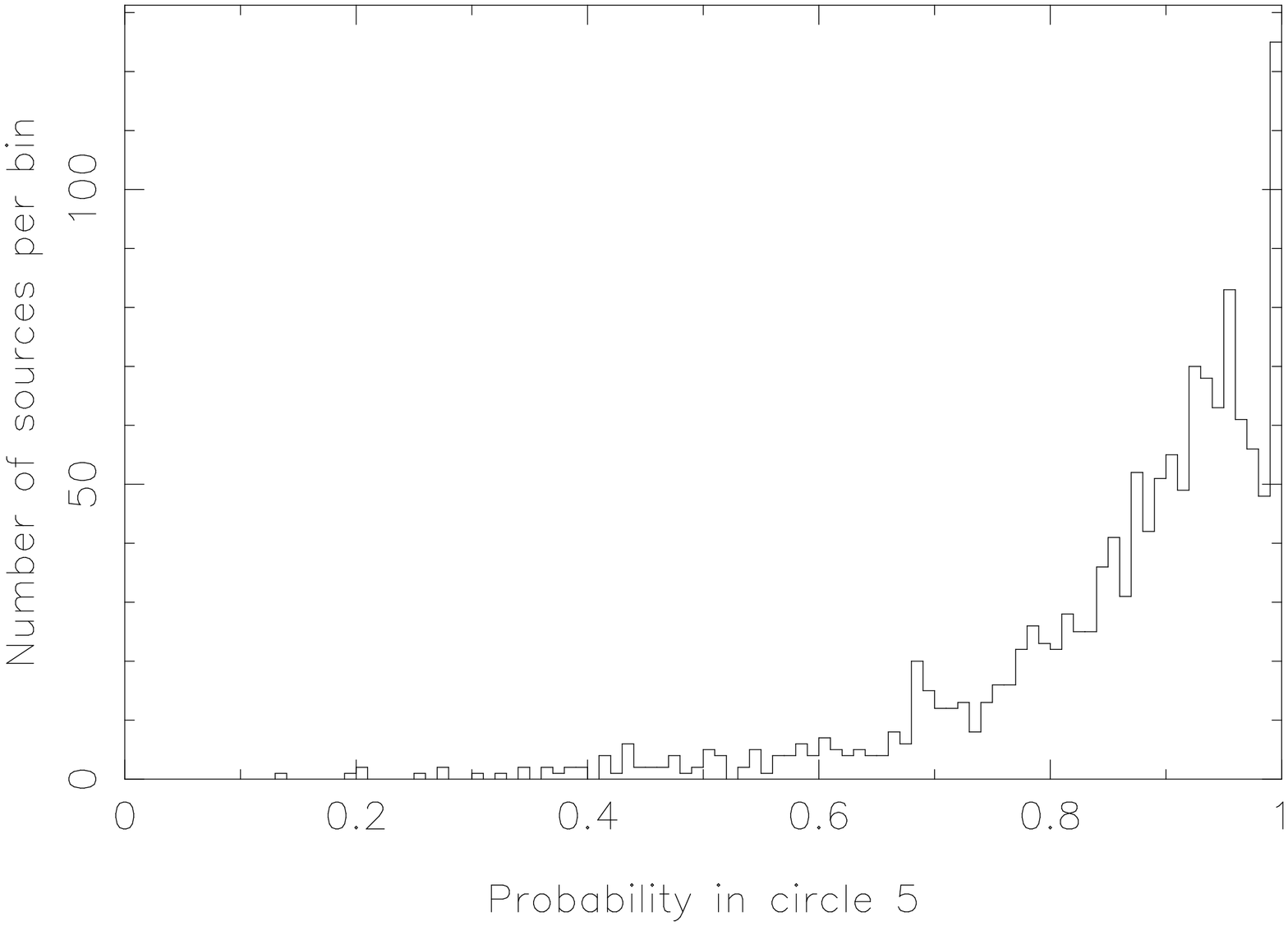,width=4.0in,angle=270}}
\caption{Galaxy concentrations yielding a high value of the statistic $P$ for
CSEARCH circles of small radius yield lower values when larger radii ($R$) are 
used, and so produce a secondary peak in the P-distribution. The plot shows this
distribution for a circle of $R=10$ arcmin, taking as a sample SGP sources
obtained using ${\rm R}=3$ arcmin and for which $0.95<{\rm P}<1$.}
\label{f5}
\end{figure*}

Second, whereas the histograms for circles 1 and 2 decay rapidly from the peak
at ${\rm P}\sim 1$ to a flat plateau, those for circles 3 and 5 decay even more 
rapidly to a minimum and then rise to a second enhancement in $P$, covering the
range $0.8<{\rm P}<0.97$ for circle 3 and $0.7<{\rm }P<0.9$ for circle 5. To 
search for the 
cause of this second enhancement we examined how candidates found using circles 
1 and 2 were registered using circles 3 through 5. A typical result 
(Figure \ref{f5}) shows how compact groupings of galaxies found by circle 2, 
for which $0.95<{\rm P}<1$, are detected using circle 5.  The number of
sources falls sharply from $P=1$ to a minimum, rises to a peak at 
${\rm P}\sim 0.92$ 
and then falls again gradually, i.e. much of this second enhancement is due to
relatively compact groups, for which the number of galaxies counted is not 
significantly enhanced when detected with circles 3 and 5. As a consequence
the effect of increasing the circle radius is to reduce ${\rm P}$.

The five candidate lists (${\rm R}$ = 1.5, 3, 5, 7.5 and 10 arcmin) are
cross-correlated to remove redundant detections, and so this second enhancement
is in principle not a problem. However, if for each circle ${\rm P}_{t}$ is set
so as to yield a high value of $C{_o}$ (equation 2), then for circles 3 through
5 the contamination ${\rm C}_{t}$ is increased substantially, making the 
subsequent screening of the sample more difficult and time-consuming. Therefore
as a precaution, for circles {\bf 4 and 5 only} we have set ${\rm P}_{t}$ at 
$0.95$, so as to include only the sharp peak. We show below that this is a 
reliable procedure. 

The SGP field contains 11981 RASS-2 sources having a likelihood greater 
than or equal to $7$. Table 1 shows the results of analysing this sample with 
CSEARCH, and summarises the assumptions made in selecting cluster candidates.
The last entry ${\rm N}_{\rm new}$ gives the yield of new candidates as one 
starts with circle 1, and then proceeds from circles 2 to 5, {\bf removing 
redundant detections}
at each stage. The trends in these numbers confirm the expectation that the 
smaller circles should detect the majority of clusters. The decision to set 
${\rm P}_{t}$ to $0.95$ for circles 4 and 5 was supported later by a search of 
the final list of 186 cluster candidates, which revealed no case in which the 
CSEARCH 
detection relied solely on circle 4 and/or circle 5. The total sample extracted 
using CSEARCH comprised 3236 candidates. Table 1 indicates that the majority of
clusters are detected using circles 1, 2 and 3, and we deduce from column 6
that the contamination of the sample should lie between 46 and 56 percent.\\

\begin{deluxetable}{ccccccc}
\tabletypesize{\small}
\tablenum{1}
\tablecolumns{7}
\tablecaption{Summary of the CSEARCH Analysis of the RASS Sources in 
the SGP Region.}
\tablehead{\colhead{Circle} & \colhead{Circle}  & \colhead{ } & 
\colhead{Range for}  & \colhead{Completeness} & 
\colhead{Contamination} & \colhead{ }\\ 
\colhead{ } & \colhead{radius}  & \colhead{${\rm P}_{t}$} & 
\colhead{estimating} & \colhead{${\rm C}_{o}$} & \colhead{${\rm C}_{t}$} &
\colhead{$N_{new}$\tablenotemark{a}} \\
\colhead{ }  & \colhead{arc min} &  \colhead{ } &  \colhead{${\rm n}_{b}$} & 
\colhead{ }  & \colhead{ } & \colhead{ }\\
}
\startdata   
    1    &  1.5  &  0.92   & $0<P<0.85$ & 0.925 &  0.46   & 1857 \\
    2    &  3    &  0.90   & $0<P<0.85$ & 0.900 &  0.50   &  884 \\
    3    &  5    &  0.90   & $0<P<0.80$ & 0.700 &  0.56   &  428 \\
    4    &  7.5  &  0.95\tablenotemark{b} & $0<P<0.70$ &   -   & $>0.75$ &
       33 \\
    5    & 10    &  0.95\tablenotemark{b} & $0<P<0.70$ &   -   & $>0.75$ &
       34 \\
         &       &         &            &       &         &      \\
\enddata 
\tablenotetext{a}{After removal of sources found using circles of smaller 
radii.}
\tablenotetext{b}{Probability threshold set to include only the sharp peak in 
the P-distribution (see text). The completeness and contamination are difficult
to estimate due to the gradient in the P-distribution.}
\end{deluxetable}

\subsection*{3.4 Reanalysis of the X-ray Data using the Growth
Curve Analysis (GCA)}
Previous studies found that the energy flux from
extended extended sources is underestimated by the standard RASS source 
detection
algorithm (Ebeling et al. 1996, 1998 and De Grandi et al. 1997), and section 2
has stated the various analyses developed to correct this problem. The 
growth curve
analysis (GCA) method, described in detail by B\"ohringer et al. (2000), has
been used in this study to analyse the RASS-2 data base. The GCA returns  
among other information the following important parameters:
background subtracted source count-rate and its 
Poisson (photon statistical) error, a locally redetermined position of the 
centroid of the source X-ray emission,
the mean exposure for the source region, a significance of the source 
detection, the radius out to which source emission
is significantly detected, a
hardness ratio characterizing the source spectrum and
its photon statistical error, a fitted source core radius,
and the probability, obtained using a Kolmogorov-Smirnov test, that the object
is consistent with being a point source. The source position is determined with 
an error
which in most cases is less than $1\ {\rm arcmin}$, the exceptions being nearby
clusters with irregular emission structure. Errors in count-rate and hardness 
ratio are given in our catalog (Table 3). Taking A2726 as an example, the hard 
band count-rate is $0.103\pm 0.022\,{\rm ct}\,{\rm s}^{-1}$ and the hardness 
ratio is $0.43\pm 0.27$.

The basic parameters are derived from the photon distribution
using the three energy bands `hard' (0.5 to 2.0 keV, channel 52 - 201),
`broad' (0.1 to 2.4 keV, channel 11 - 240), and `soft' (0.1 to 0.4
keV, channel 11 - 40). The band definitions are the same as those used
in the standard analysis (Voges et al. 1999).
Most of the derived parameters are based on analyses of data in
the hard energy band, as clusters are detected
in this energy band with the highest signal-to-noise ratio.
One of the exceptions is the hardness ratio, which 
requires also the results from the soft band. 

\begin{figure*}
\centerline{\psfig{figure=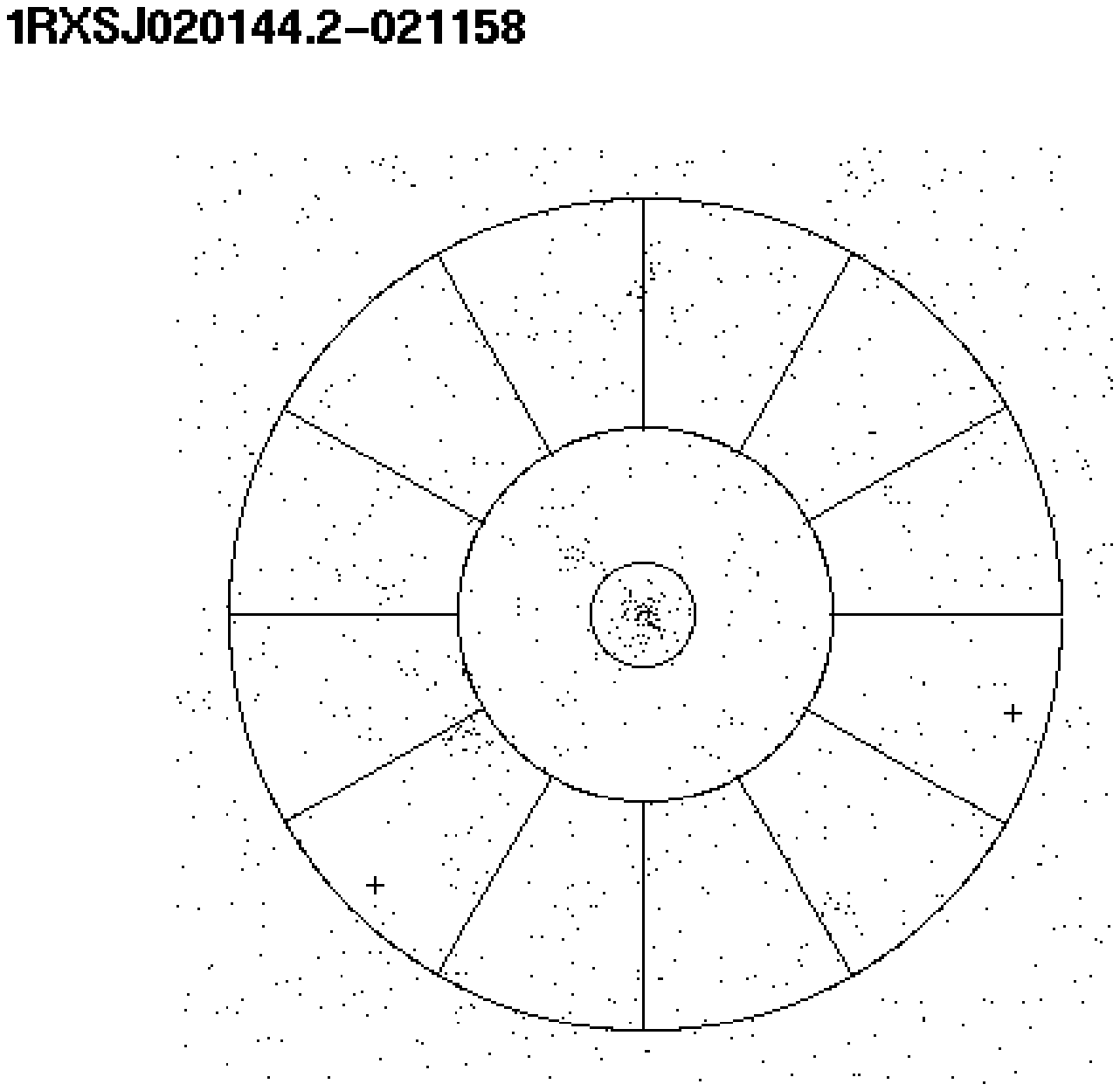,width=4.0in,angle=0}}
\caption{The configuration used in the growth curve analysis (GCA) of a ROSAT
source. The image shows the distribution of photon events in the hard band
($0.5-2.0\,{\rm keV}$) within a square field of width $1.5\,{\rm deg}$, centered
on the 
source to be studied. The background is measured in the annulus between the
two outer circles of radii 20 and 41.3 arcmin, which is divided into 12 
sectors.
A sector is discarded if its background has too high a deviation from the
average. Two examples are marked with a cross. The small inner circle 
designates the radius out to which
significant X-ray emission was detected by the GCA for this source.}
\label{f6}
\end{figure*}

The GCA method uses the photon event distributions in two circles 
(Figure \ref{f6}), of radii 20 and 41.3 arcmin, centered initially on 
the RASS-2 source center. Events within the inner circle are analysed to 
determine the source characteristics, and those in the outer annulus are used
to determine the background level. In some cases special analyses were 
necessary, where either clusters had an unusually large angular extent, or 
where two or more sources lay in a confused region, requiring the use of 
'deblending' algorithms.  

\begin{figure*}
\centerline{\psfig{figure=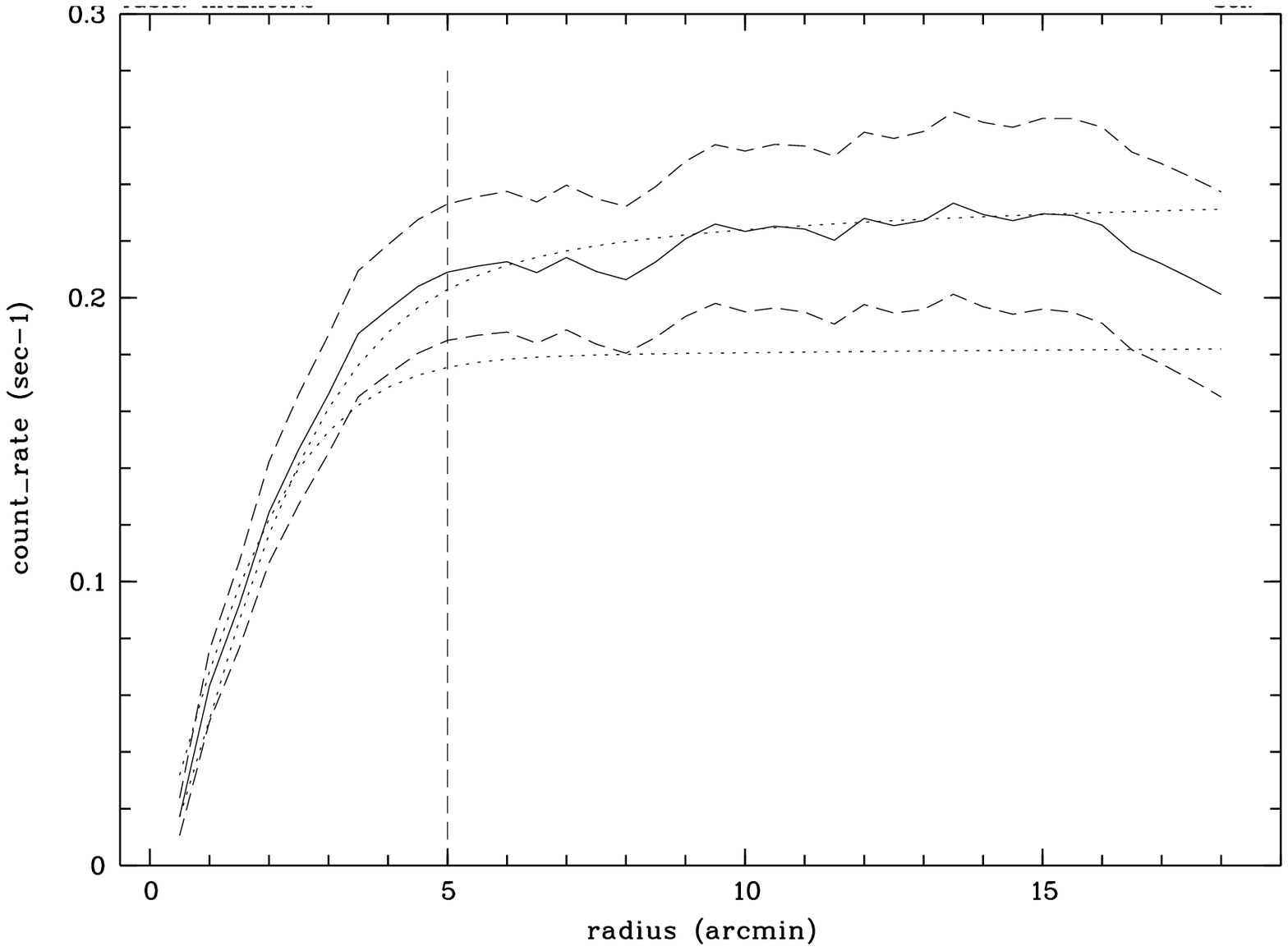,width=4.0in,angle=0}}
\caption{Integrated count-rate profile for the source shown in 
Figure \ref{f6}. The integrated count-rate profile is 
background-subtracted and
the two dashed curves give the limits determined by the $1{\rm \sigma }$ 
statistical
error, which includes the uncertainties of the signal and the background
determination. The vertical dashed line shows the outer source radius, as
defined in the text. The lower dotted curve shows the ${\rm \chi }^{2}$ fit of 
a point source to the data, and the upper one shows the best King-model fit.}
\label{f7}
\end{figure*}

The source count-rate is determined from the cumulative, radial source count 
rate profile ('growth curve') after background subtraction. The construction 
of the growth curve is preceded by a redetermination of the source center and 
by the background measurement. A typical example for the resulting growth
curve is displayed in Figure \ref{f7}, showing the count-rate plotted 
as a function of integration radius (solid curve). The dashed lines show the
limits determined by applying the photon event statistical error (including the
error for the background subtraction).

The count-rate is determined in two alternative ways. In the first
a significant outer radius of the source is defined
as the point where the increase in the $1{\rm \sigma }$ error is larger
than the increase of the source signal. The integrated count-rate 
is calculated using events within this 'cutoff' radius. In the second 
method a horizontal level is fitted to the outer region of the plateau, and this
value is adopted as the standard in quoting hard-band X-ray count-rates and in
calculating cluster energy flux and luminosity (section 4). However, we use
the first method also as a check on the count-rate and as a means of estimating
systematic uncertainties in the count-rate
determination, which must be combined with the pure photon statistical errors.

Careful examination of the overlays of X-ray and optical images (section 3.5.1)
of the cluster sample revealed 19 cases in which the source was very extended, 
or source confusion might cause errors in deriving the source count-rate. 
These 19 were subjected to the special analyses referred to above, in
which the source count-rates were revised. Six systems were identified 
as double clusters and not subjected to the deblending analysis. They are
identified as double clusters in the survey catalog (Table 3).

The other two important source properties determined by the GCA method are the
spectral hardness ratio and the source extent.
The hardness ratio, $\rm HR$, is defined as
${\rm HR} = {{\rm H} - {\rm S} \over {\rm H} + {\rm S}}$
where ${\rm H}$ is the hard band and ${\rm S}$ the soft band source count-rate
(both determined for the same outer radius limit).

The source extent is investigated in two ways.
In the first a ${\rm \beta }$-model profile (Cavaliere \&  Fusco-Femiano 1976)
convolved with the averaged survey PSF is fitted to the differential
count-rate profile (using a fixed value of ${\rm \beta =\,2/3}$), and we 
derive from the results a core radius, ${\rm r}_{c}$. In the second a
Kolmogorov-Smirnov (KS) test is used to estimate the probability that
the object is a point source. It is taken as highly likely that the source is 
extended when the KS probability is less than
0.01. Tests with X-ray sources which have been identified with
stars and AGN showed that a false classification as an extended source
occurred in less than 5\% of the sample.

The final step in the X-ray analysis was to establish a lower limit to the 
source count-rate, in order to set practical limits to source identification 
and redshift determination. The list of 3236 candidates from the CSEARCH 
analysis (section 3.3) was truncated using a threshold of 
$0.08\,{\rm ct}\,{\rm s}^{-1}$ for the cut-off radius count-rate in the hard 
band. This reduced the list of candidates to 477. A sharp limit in count-rate 
corresponds to a range of energy flux, determined by the variation of
interstellar column density $({\rm N}_{H})$ in the field, which leads to some
incompleteness at low fluxes. B\"{o}hringer et al. (2001) estimate a range 
of $1.55-1.95\,10^{-12}\,{\rm erg}\,{\rm s}^{-1}$ in the band
$0.1-2.4\,{\rm keV}$ for $10^{20}<{\rm N}_{H}<10^{21}\,{\rm cm}^{-2}$. In the 
SGP field ${\rm N}_{H}$ takes on relatively low values so that, for example, 
for only 8 percent of the clusters selected is 
${\rm N}_{H}>4.0\times 10^{20}\,{\rm cm}^{-2}$.
Therefore we can be certain that this source of incompleteness is absent 
at fluxes greater than $2\,10^{-12}\,{\rm erg}\,{\rm cm}^{-2}\,{\rm s}^{-1}$.

\subsection*{3.5 Removal of Contaminating Sources from the
Cluster Candidate List.}
The removal of contaminating sources from the list of 477 candidates was by 
necessity an iterative process, in which the decisions made were reviewed 
repeatedly to ensure that the criteria used had been applied consistently. The 
following sections describe the steps in this process.
\subsection*{3.5.1 Examination of COSMOS Finding Charts and 
Overlays of X-ray and Optical Fields.}
\begin{figure*}
\centerline{\psfig{figure=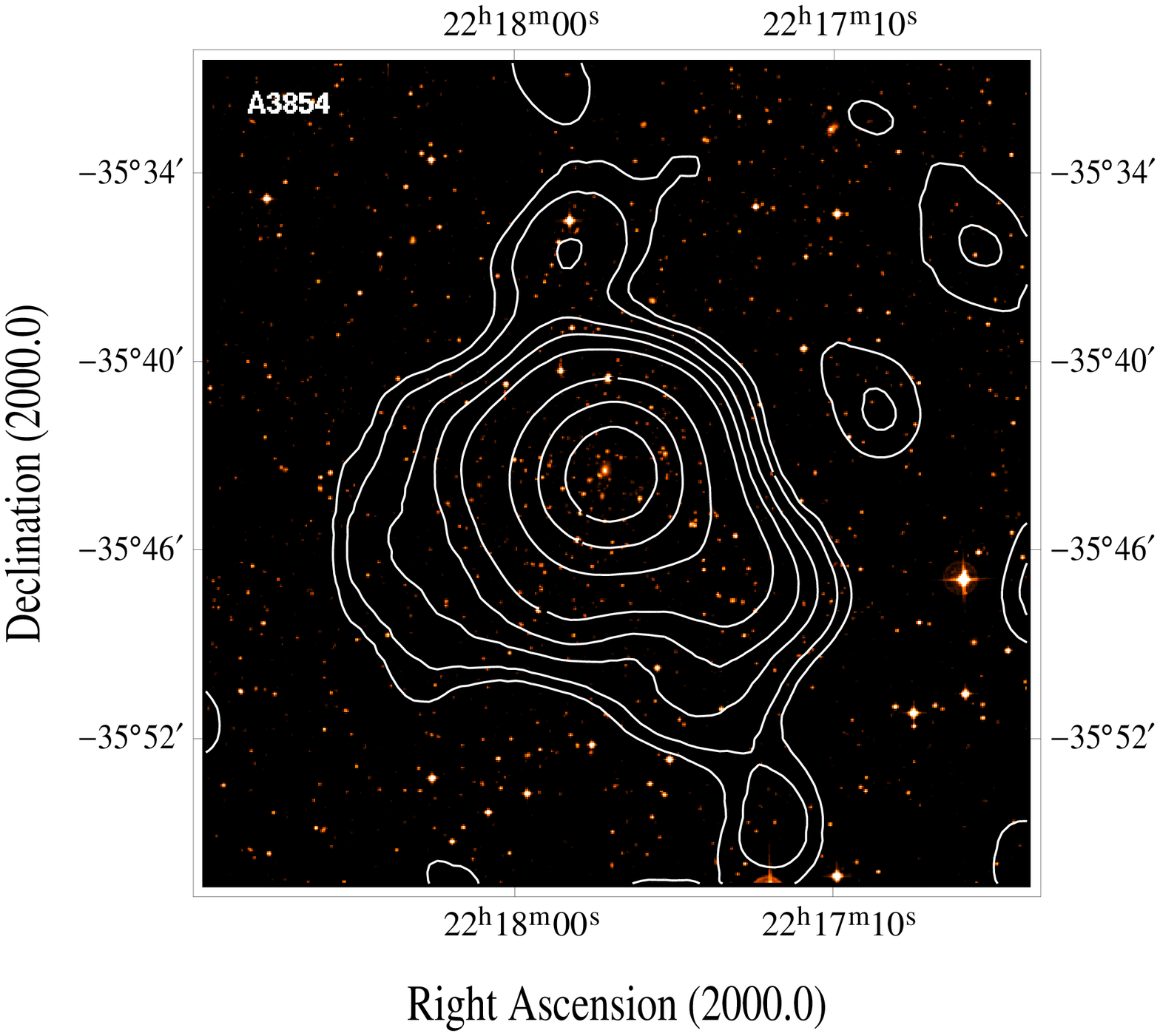,width=4.0in,angle=0}}
\caption{A contour plot of the X-ray flux from A3854 in the hard band 
($0.5-2\,{\rm keV}$), superposed on the COSMOS digitised image obtained from the
UK Schmidt IIIa-J survey}
\label{f8}
\end{figure*}

For each cluster candidate a finding chart $7\times 7\,{\rm arcmin}^{2}$ in 
size was extracted from the COSMOS data base. In addition an overlay was made 
of the X-ray surface brightness contours upon the COSMOS optical 
field. For example Figure \ref{f8} shows the overlay made for the 
Abell cluster A3854. 
Examination of these finding charts and overlays during the selection process
has been valuable for several reasons.
First, visual scrutiny has removed cases in which CSEARCH had been 
deceived by artifacts, for example satellite trails, the haloes and 
diffraction spikes of bright stars, and the attempt by COSMOS to resolve the
structure of bright nearby galaxies. These structures often were resolved by 
COSMOS into objects identified as galaxies. Second, the RASS-2 analysis 
resulted in multiple detections of extended sources such as clusters, which 
may be removed after examination of the X-ray/optical overlays. As described 
in section 3.3, a significant number of redundant detections was removed to 
minimise
any bias of the CSEARCH analysis. However this process was not perfect, and it 
was possible later to identify further cases by careful re-examination of the 
overlays. The latter are counted in the detection statistics shown later in 
Table 2. Finally, scrutiny of the finding charts and overlays has identified 
occasional errors, which are simple to correct, where a CSEARCH candidate 
is identified with a faint RASS-2 X-ray source having a brighter X-ray 
neighbour. The GCA technique then has preferred the latter and derived the 
higher count-rate. This problem was detected through visual examination of the 
overlays.

\subsection*{3.5.2 Rejection Criteria based on X-Ray Hardness 
Ratio and Angular Extent.} 
The temperature and extent of the hot gas in clusters of galaxies make
it possible to set limits to both the X-ray hardness ratio and the angular
extent of a RASS cluster source (Ebeling et al. 1996, 1998). This provides 
effective discriminants between 
clusters and the majority of contaminating sources in the CSEARCH sample, made
up of AGN and stars. In this section we justify the following two
criteria, which have been used in removing a large fraction of these 
contaminating sources:\\
\hspace*{1cm}1) The {\bf upper limit} of the X-ray hardness ratio (section 3.4)
is less than zero, \\
\hspace*{1.5cm}i.e. if ${\rm HR}_{ul}<0$, the source is too soft.\\
\hspace*{1cm}2) The core radius ${\rm r}_{c}$ obtained from the angular extent 
calculation is less than \\
\hspace*{1.5cm}$0.5\,{\rm arcmin}$, i.e. the object is a point source.\\ 
These criteria were applied independently in rejecting sources. In order to 
establish criteria using these parameters, we selected control 
samples describing the characteristics of clusters, AGN and stars. The first 
contained $118$ Abell clusters extracted from the total sample of $477$ 
candidates (section 3.4),
using as a criterion a correlation between the X-ray position and the centroid 
of an Abell cluster. Conservatively we excluded from the sample clusters where
the separation between the ROSAT and Abell positions was greater than 
$5\,{\rm arcmin}$, where the survey exposure was low $(<100\,{\rm s})$, or 
where X-ray 
maps revealed nearby components in the source region, one of which might 
not be a cluster. In each case the X-ray/optical overlay image (section 3.5.1)
and the COSMOS finding chart were examined. The count-rates of these clusters
covered the full range of the study reported here.

The second and third control samples comprised AGN and stars detected by the 
ROSAT survey and identified through spectroscopic optical observations. This 
project was a collaboration between MPE and ESO, in which the error circles of 
all RASS-1 sources in four relatively small regions of the southern hemisphere 
were searched using
the COSMOS data base to identify optical candidates. The nature of these 
candidates was examined through spectroscopic observations, after which the 
likely source of the X-ray emission was identified (T. A. Fleming 1994, private
communication). Using 
three of these fields having a total area of $570\,{\rm sq}\,{\rm deg}$, we 
selected samples
of $138$ AGN and $61$ stars, whose positions are within 20 arcsec of the RASS-2
source positions. Again, the X-ray count-rate of these sources covered the full
range of the cluster survey reported here. 

\begin{figure*}
\centerline{\psfig{figure=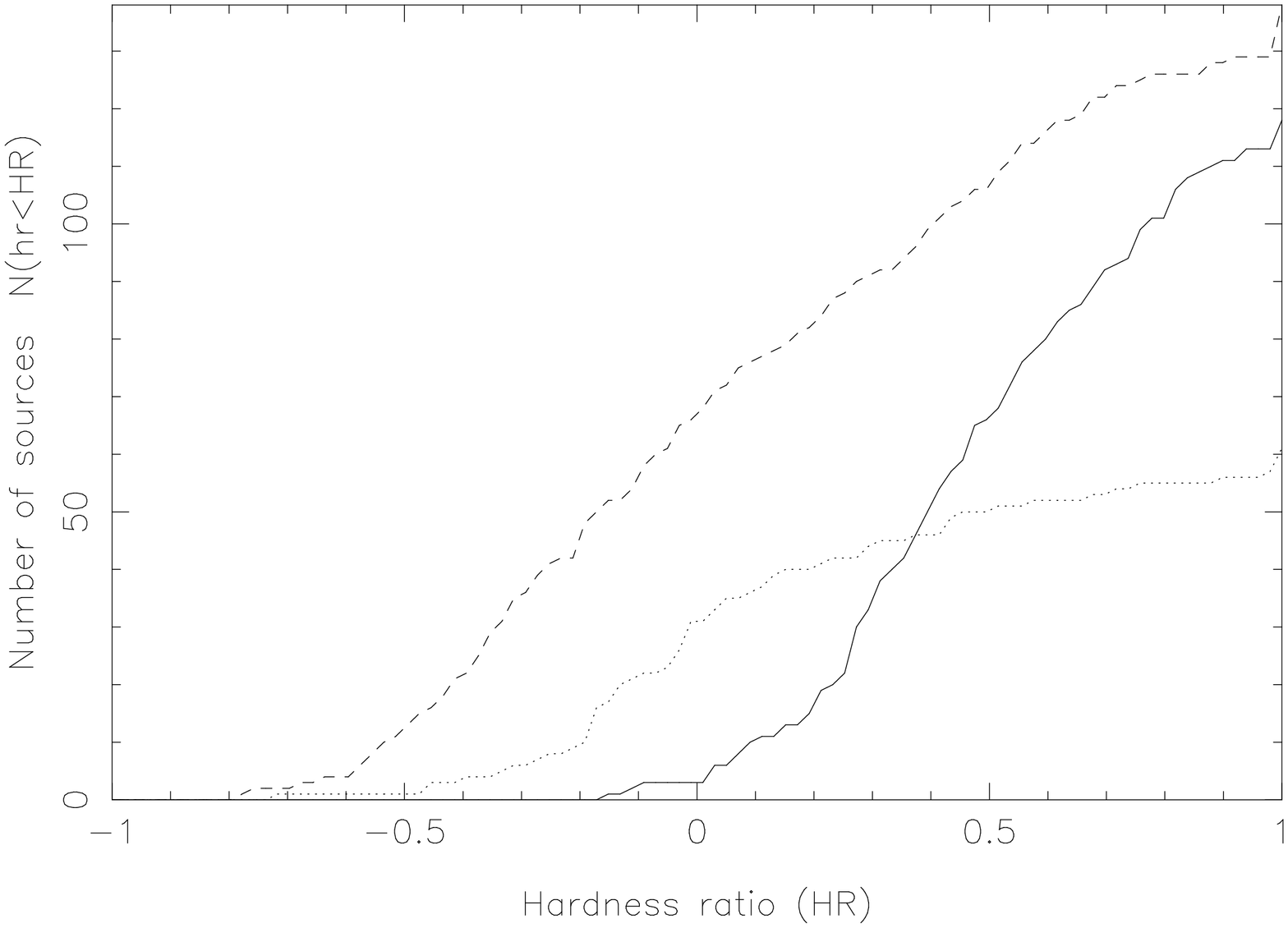,width=4.0in,angle=270}}
\caption{The cumulative distributions of hardness ratio for selected samples of
 ROSAT sources are shown by the dashed line for AGN and by the dotted line for 
 stars. The samples comprise all RASS sources in three special fields selected
 for an ESO Key Project (not the Key Project described in the text in regard to
 cluster redshift determination), which were identified by spectroscopic optical
 observations. The solid line is the distribution for an unambiguously 
 identified sample of Abell clusters of galaxies in the SGP region.}
\label{f9}
\end{figure*}

The hardness ratio (${\rm HR}$) distributions for these samples are shown in 
Figure \ref{f9}. Whereas ${\rm HR}>0$ for $97$ percent of the Abell clusters 
in our test sample, 
${\rm HR}<0$ for $53$ and $51$ percent of the AGN and star samples 
respectively. 
Therefore a conservative threshold of zero for the {\bf upper limit} to 
${\rm HR}$ has
been set, above which RASS sources qualify for retention in the list of cluster 
candidates. This is an effective method of removing a large fraction of the
stars and AGN in our sample of 477 candidates. 

\begin{figure*}
\centerline{\psfig{figure=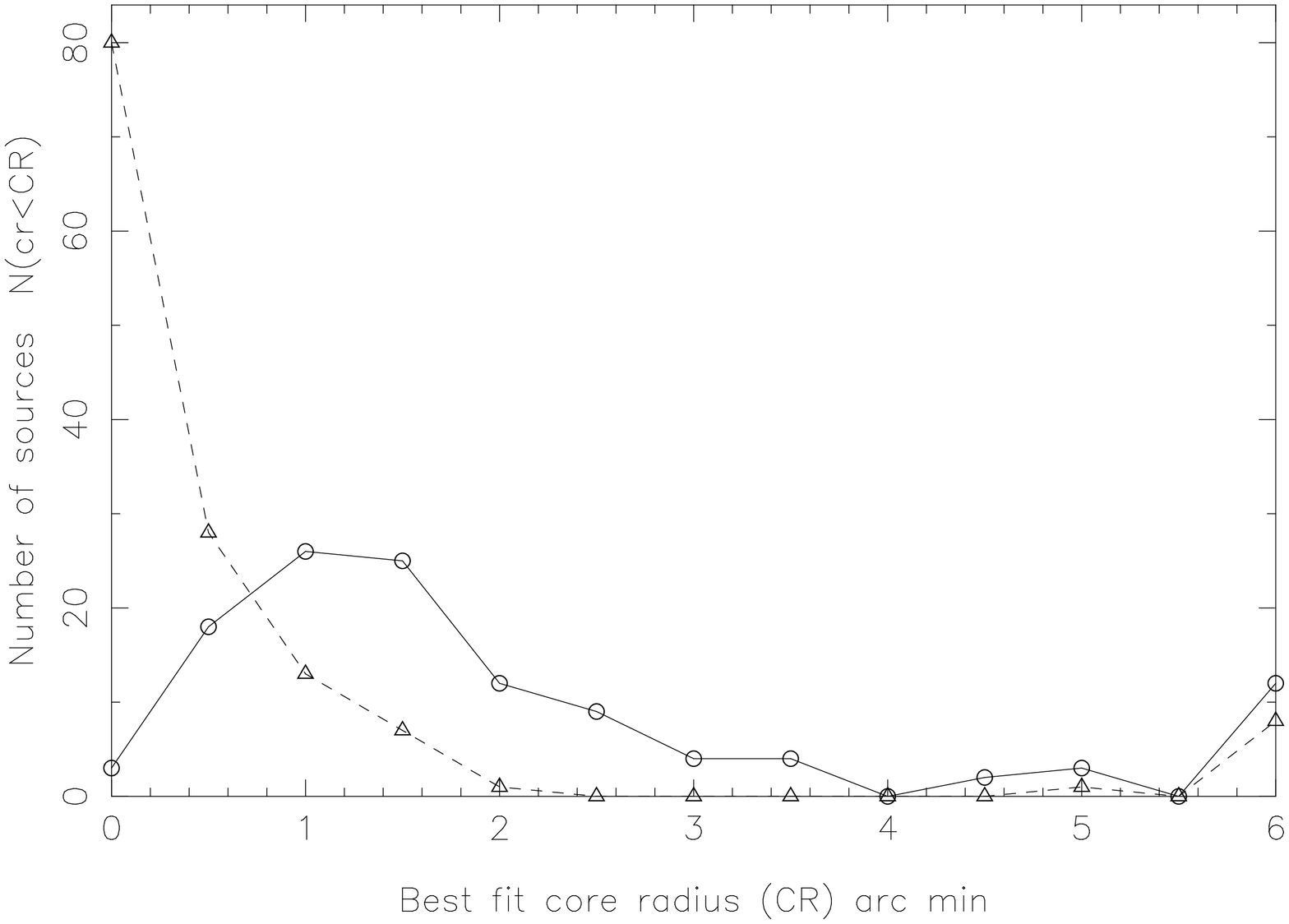,width=4.0in,angle=270}}
\caption{The distributions of core radius obtained by fitting a King model for
 cluster surface brightness distribution to the samples of AGN and clusters
 used in deriving Figure \ref{f9}. AGN are represented by the dashed 
 line and clusters of galaxies by the solid line. The rise in the distributions
 near a core radius of 6 arcmin is discussed in the text.}
\label{f10}
\end{figure*}

The angular extent, as measured by the derived core radius ${\rm r}_{c}$, 
provides 
an effective test for distinguishing clusters from AGN, which however runs into 
two difficulties. The first is caused by the diminishing size of clusters with 
increasing redshift, an effect which is emphasised if the cluster contains a 
cooling flow. The second is the effect of source confusion in a RASS-2 
detection of an AGN, when the GCA analysis may yield a finite value for the 
source extent. These difficulties are constrained to a certain extent by the 
redshift limit (${\rm z}\sim 0.3$) and by the limit set to the X-ray count-rate
in the hard band (section 3.4). The distributions of ${\rm r}_{c}$ derived from
the two samples are shown in Figure \ref{f10}. The rise in these curves at a 
radius of $6\,{\rm arcmin}$ is a result of setting a limit 
${\rm r}_{c}\leq \,6\,{\rm arcmin}$ in the GCA, and to a lesser extent of the
influence of surrounding sources on the growth curve analysis (section 3.4). 
The important characteristic is that  
58 percent of AGNs satisfy ${\rm r}_{c}=0$, whereas $97$ percent of 
Abell clusters in our test sample satisfy the condition 
${\rm r}_{c}\geq 0.5\,{\rm arcmin}$. 
Therefore only those sources among the 477 CSEARCH cluster candidates were 
retained which satisfied the condition ${\rm r}_{c}\geq 0.5\,{\rm arcmin}$. As 
discussed in
section 3.5.5, this results in some incompleteness near the survey limit, where
the RASS and COSMOS data become marginal in the detection of distant clusters. 

We have examined four clusters in the Abell control sample, which were rejected
when these
hardness ratio and core radius criteria were used. One, associated with A2800,
failed the hardness ratio test. This may be due to contamination, as it is in a
confused region of X-ray emission. Three sources, associated with A3866, S181 
and S1121,
failed the core radius test. In the case of S181 the galaxies are faint and
displaced from the center of the X-ray source, which is comparatively strong
($0.134\,{\rm ct}\,{\rm s}^{-1}$ in the hard band). The small RASS-2 error 
circle contains a
stellar object (${\rm b}_{j}=17.7$), and therefore we conclude that this source
is an AGN (the emission from AGN comes from a concentrated region, so that 
often COSMOS classifies the object as a star). The source coincident with A3866
has a high count-rate ($0.4\,{\rm ct}\,{\rm s}^{-1}$) yet shows only weak signs
of extent.
Although the spectroscopic observations (Romer 1995) revealed evidence of an 
AGN, overlaying an EINSTEIN HRI image upon an optical DSS image indicates
that the central object is probably a cD galaxy and that the X-ray source is 
very compact but not a point source. Thus the decision to reject this source 
is open to correction. S1121 appears to be an example of a distant cluster
near the limit of the survey sensitivity. Thus four clusters in the test sample
of 118 Abell clusters were rejected, from which we estimate that in applying 
the hardness ratio and core radius criteria to our cluster sample, the loss of
clusters should be less than 4 percent.
 
\subsection*{3.5.3 Removal of Bright Stars.}
Bright stars appear in the sample of cluster candidates either because
of a chance association with a region of high galaxy density, or because
COSMOS has broken down the diffraction spikes and/or halo into objects
classified as galaxies. They have been removed using two tests, one being the
hardness ratio test described in section 3.5.2 and one being a statistical test.
In the latter we have analysed $17$ regions of diameter $1^{\rm o}$ in the SGP 
field, using the COSMOS data to derive the mean density of stars as a function 
of ${\rm b}_{j}$ magnitude. The following argument was used to reject RASS-2 
sources having stars brighter than ${\rm b}_{j}=12$ in the error circle. The 
density of stars brighter than ${\rm b}_{j}=12$ was found to be 
$0.0165\,{\rm arcmin}^{-2}$, 
which implies that the probability of finding such a star by chance in a RASS-2 
90-percent confidence error circle, which typically has a radius of about 
$30\,{\rm arcsec}$, is $0.013$. Therefore if we reject all candidates having a 
star brighter than ${\rm b}_{j}=12$ within the error circle, the loss of 
clusters from the sample should be no more than $1-2\,{\rm percent}$. 

This procedure has been applied by correlating the candidate list with stars in 
the SIMBAD and TYCHO (H{\o}g et al. 1998) catalogs. A correlation of the 
${\rm b}$ 
magnitudes in these catalogs with the COSMOS ${\rm b}_{j}$ magnitudes shows 
that the latter are lower by about 0.5 magnitudes for stars fainter than 
${\rm b}_{j}\sim 10$, but that the difference widens steadily for brighter 
stars. Therefore in examining bright SIMBAD or TYCHO catalog stars in the 
RASS-2 error circle we set a threshold of ${\rm b}=12.5$, below which the star 
was judged to be the cause of the X-ray emission.

Additional stars, not listed in the SIMBAD and TYCHO catalogs, were 
identified later by inspection of the COSMOS finding charts and the 
X-ray/optical overlay images for the 477 candidates. 
\subsection*{3.5.4 Removal of AGN and Large Extended Emission Regions.}
The X-ray sources removed by the hardness ratio $({\rm HR}_{ul}<0)$ and
core radius $({\rm r}_{c}<0.5\,{\rm arcmin})$ tests have been reviewed 
carefully, 
using correlations of the list of 477 candidates with the Veron Catalog of 
active galaxies (Veron-Cetty \&  Veron 1998) and the NASA Extragalactic Data 
Base (NED) to identify sources for which the 90 percent RASS-2 error circle 
contained an identified AGN, and examination of the X-ray/optical overlay
images and the COSMOS finding charts. However in the sample remaining after
these tests, after removal of stars (section 3.5.3) and after identification of
detections
resulting from COSMOS artifacts, there remained a contaminant comprising 28
sources for which ${\rm HR}_{ul}\geq 0$ or ${\rm r}_{c}\geq 0.5\,{\rm arcmin}$,
but whose identification as a cluster was suspect. These sources
were reviewed using again the Veron and NED correlations, overlay images and 
COSMOS finding charts, but in addition using the results of the CSEARCH
analysis. Of these 28 we rejected 9 as being AGN and 1 as
an X-ray source superposed on a nearby galaxy over 5 arcmin in extent. Nine
sources were excluded because the extent detection was
due to source confusion, and 8 because CSEARCH either made
an identification at a large radius only, inconsistent with the small X-ray
extent, or made an error due to COSMOS artifacts. The one remaining source
requires further study and was rejected tentatively. There is a possible
association with A2813, but the minimum found by GCA in the core radius fit
was very shallow and the evidence for a cluster in the X-ray/overlay image
was indecisive.

\begin{figure*}
\centerline{\psfig{figure=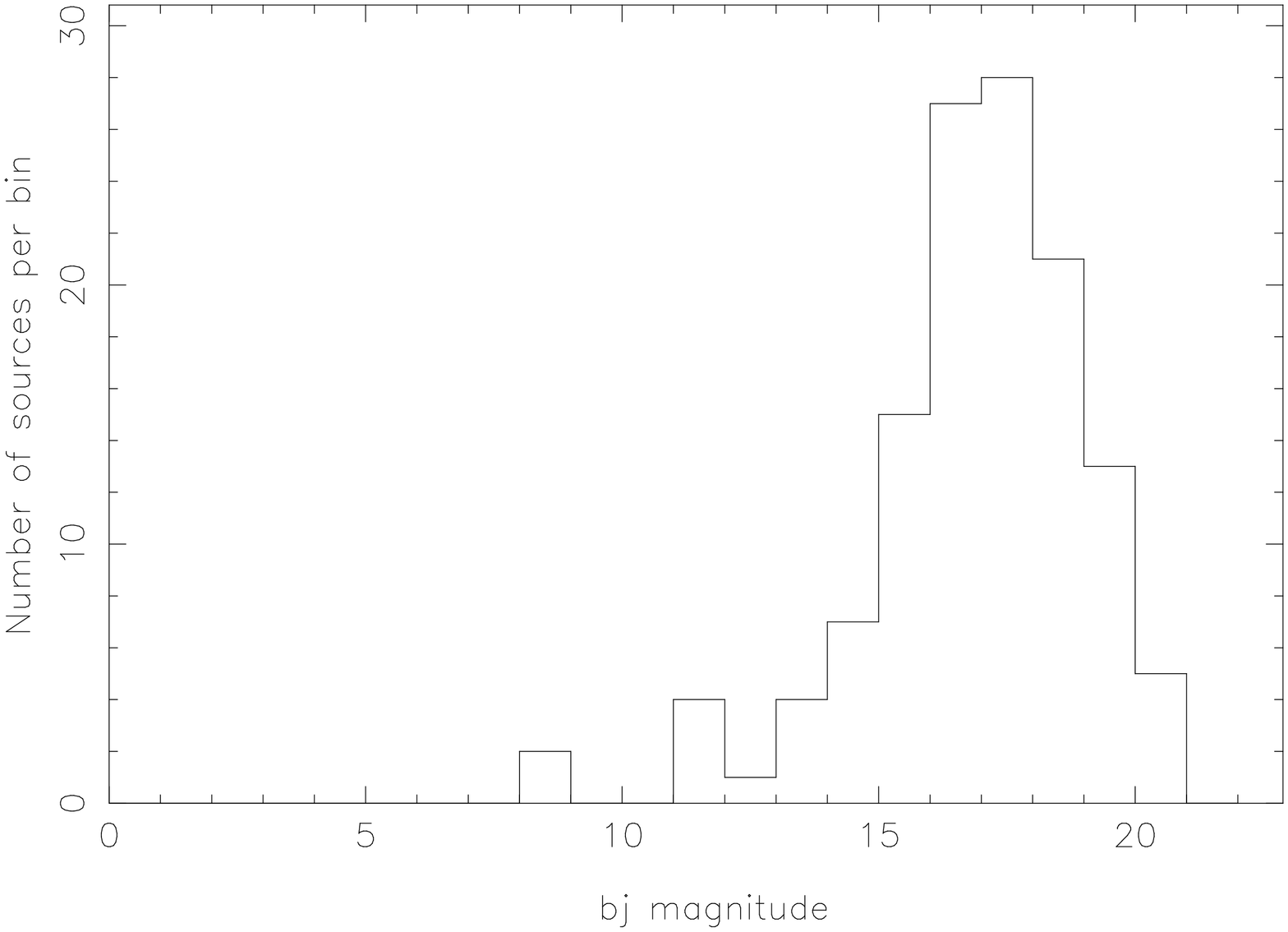,width=4.0in,angle=270}}
\caption{The distribution of ${\rm b}_{j}$ magnitude for the sample of sources
identified in the SGP region as AGN.}
\label{f11}
\end{figure*}

The total number of AGN identified was 127, and Figure \ref{f11} 
shows the distribution of their ${\rm b}_{j}$ magnitudes, as measured by 
COSMOS. Where the AGN was not identified in the Veron catalog or NED, 
${\rm b}_{j}$ for the
brightest object inside the 90 percent error circle was used. The half-width of
the distribution lies in the range $15<{\rm b}_{j}<19$. 

Finally, we discovered five candidates showing large extended regions of X-ray 
emission $10-20\,{\rm arcmin}$ in diameter, whose optical fields show no signs 
of a significant galaxy population and whose origin as the confusion of several 
sources is not obvious.

\subsection*{3.5.5 Summary of the Selection 
Statistics and Estimate of the Sample Completeness.}
After removal of RASS-2 sources contaminating the SGP sample of 477 cluster
candidates, using the procedures described above, there remained 186 candidates.
The statistics of this selection process are summarised in Table 2.
Of the 61 redundant RASS-2 detections in Table 2 which eluded the 
earlier screening, 24 were associated with clusters. Therefore of
the 477 candidate sources ${\rm N}_{eff}= 210$ are associated with 
clusters, which implies that the contamination of the sample of 477 was 0.56. 
This is in marginal agreement with the results of the CSEARCH analysis 
(section 3.3), which predicted $0.46<{\rm C}_{t}<0.56$. Examination of Table 1 
shows that $84.7\,{\rm percent}$ of the cluster candidates were identified by 
CSEARCH using the
smaller circles 1 and 2, and consequently that formally the contamination of the
sample should be 0.5 or less, and the completeness about 0.90. Therefore we can
estimate that the sample of 477 candidates should contain $0.5\times 477=238$ 
clusters and that a complete sample obtained using CSEARCH would have yielded 
${\rm N}_{tot}=0.5\times 477/0.9=265$ clusters. 
Therefore we can estimate that the overall completeness of the sample of 
selected clusters is ${\rm N}_{eff}/{\rm N}_{tot}=210/265=0.79$. This
incompleteness, which is the sum of the three effects summarised in section 2, 
contributes to the flattening of the log N/log S distribution at fluxes below 
$3.0\times 10^{-12}\,{\rm erg}\,{\rm cm}^{-2}\,{\rm s}^{-1}$ in the band 
$0.1-2.4\,{\rm keV}$ (section 5).

\begin{deluxetable}{cc}
\tabletypesize{\small}
\tablenum{2}
\tablecolumns{2}
\tablecaption{Summary of the Results of Analysing the Sample of RASS-2
Cluster Candidates in the SGP Region.}
\tablehead{  
\colhead{Source} & \colhead{Number}
}
\startdata 
    Clusters of galaxies                     & 186 \\
    Active galactic nuclei (AGN)             & 127 \\
    Stars                                    & 74 \\
    Extended X-ray emission regions          &  5 \\
    Redundant RASS-2 detections              & 61 \\
    Sources below X-ray count-rate threshold & 24 \\
                                                   \\
    Total                                    & 477 \\
\enddata 
\end{deluxetable}

The above argument implies that $238-210=28$ clusters have been missed in 
searching the 477
candidates. This has arisen in most cases while applying the criteria
${\rm HR}_{ul}<0$ and ${\rm r}_{c}=0$ to reject candidates, and in section 
3.5.2 we have estimated that applying these criteria applied to a control 
sample of Abell clusters should yield an incompleteness of $4-5\,{\rm percent}$.
Two examples of clusters with overly soft spectra are A2800 
(${\rm HR}=-0.15\pm 0.11$) and the cluster $2310-73$ (${\rm HR}=-0.19\pm 0.13$)
discovered by Tucker et al. (1995) in the EMSS. In neither case is the nature 
of the contaminating source clear.

One limit of this study is that the criterion 
${\rm r}_{c}\geq 0.5\,{\rm arcmin}$ for 
selecting cluster candidates will inevitably discriminate against the more
distant clusters of small angular extent, particularly those in which a cooling 
flow makes the X-ray source more compact. This loss is evident in the
flattening of the log N-log S distribution at low flux (section 5). Three 
aspects of the analysis bear out this trend. First, of the sources rejected 
because ${\rm r}_{c}=0$ we identified at least 4 cases, where the X-ray/optical
overlays 
showed suggestive but marginal evidence that they were clusters. Second, there 
were 11 in our catalog of 186 clusters, for which the detection of angular
extent appeared to be near its limits. The GCA yielded a small but non-zero 
core radius, but simultaneously the KS-test yielded significant probabilities, 
between $10$ and $50$ percent, of the source being point-like. Third, in
the case of distant clusters accepted by CSEARCH, COSMOS classified some of the
galaxies as 'faint objects', which the CSEARCH analysis did not count. This is 
obvious in some
of the X-ray/optical overlays of these objects, which show the dominant
elliptical galaxy with halo surrounded by faint objects. Therefore we conclude
that some distant clusters may have been missed by CSEARCH.

\section*{4. CLUSTER X-RAY ENERGY FLUX AND LUMINOSITY}
In order to convert the ROSAT count-rate into an energy flux outside the galaxy,
the spectrum of the hot intracluster medium (ICM) has been convolved with the 
response functions of the interstellar medium (ISM) and the instrument. The
plasma emissivity was derived using the version of the Raymond \& Smith (1977)
code installed in 1994 in the MPE EXSAS software. The technique has been 
verified by comparing results with those obtained using the EXSPEC analysis 
code (B\"{o}hringer et al. 2000). The 
cluster plasma was assumed to contain elements heavier than hydrogen having
abundances 0.35 times the solar abundance. The HI column 
densities used in calculating the ISM absorption have been taken from Dickey 
\&  Lockman (1990). These calculations have been made over ranges of 
temperature between 0.05 and 10 keV ($5.8\times 10^{5}$-$1.16\ 10^{8}$ K), and 
of HI column density between $10^{19}$ and $10^{22}\,{\rm cm}^{-2}$. The 
resulting 
matrix was used to convert the count-rate in the hard band to an energy flux
$f_{\rm x}$ in units of ${\rm erg}\,{\rm cm}^{-2}\,{\rm s}^{-1}$ in the 
0.1-2.4 keV band.

For each cluster whose redshift has been measured the X-ray luminosity has 
been calculated using the relations:\\
\begin{equation}
{\rm L}_{x}={\rm K}\ f_{x}\ 4\pi {\rm D}_{L}^{2}
\end{equation}
\begin{equation}
{\rm D}_{L}=\frac{c}{{\rm H}_{0}q_{0}^{2}}(q_{0}z+(q_{0}-1)((2q_{0}+1)^{1/2}-1))
\end{equation}
where ${\rm L}_{x}$ is the luminosity in the 0.1-2.4 keV band, ${\rm K}$ is the
correction for redshifting of the radiation, and ${\rm D}_{L}$ is the 
luminosity distance. K has
been calculated for ranges of redshift between 0.025 and 0.5, and of ICM
temperature between 0.05 and 10 keV. The resulting matrix was used to convert
energy flux into luminosity.

Measured cluster temperatures have been taken from the compilation in Table 1 
of White et al. (1997). Where measured temperatures have not been found we have
used the luminosity-temperature correlation derived by White et al. (1997):\\
\begin{equation}
{\rm L}_{bol}=0.0478\,{\rm T}^{2.98}
\end{equation}
where ${\rm L}_{bol}$ is the X-ray luminosity in the 0.01-80 keV band and T is 
the temperature in keV. The assumed temperature was used to calculate a 
bolometric correction to the luminosity. For each source a cycle of iterations 
was performed in the flux and luminosity calculations, until the change in 
temperature fell below 1 percent. Where neither redshift nor temperature has
been measured, we have assumed a temperature of $5\,{\rm keV}$. 
\section*{5. RESULTS OF THE SGP CLUSTER SURVEY}
The procedures described in sections 2, 3 and 4 for searching the RASS-2
data base for cluster candidates, selecting clusters from this sample and
determining their X-ray characteristics, have resulted in the catalog of 186 
clusters shown in Table 3. The following notes, listed by the column number in 
the table, augment the information given in the header to each column.\\
({\bf 1}) ROSAT X-ray source name.\\
({\bf 2}) Right ascension (h:m:s J2000) of the centroid (GCA) of the cluster 
X-ray emission.\\
({\bf 3}) Declination (d:m:s J2000) of the centroid (GCA) of the cluster
X-ray emission.\\
({\bf 4}) Count-rate in ${\rm ct}\,{\rm s}^{-1}$ in the hard band 
($0.5-2.0\,{\rm keV}$), with error (GCA).\\
({\bf 5}) X-ray hardness ratio, with error (GCA).\\
({\bf 6}) Total ROSAT survey exposure time for the cluster, in seconds.\\
({\bf 7}) Column density of interstellar atomic hydrogen at the cluster 
center, in units of $10^{20}\,{\rm cm}^{-2}$.\\
({\bf 8}) Temperature of the cluster gas, derived by the procedure 
described in section 4, in ${\rm keV}$. The superscript $'e'$ indicates 
that no measurement was found, and the temperature was determined using an 
${\rm L}_{bol}-{\rm T}$ correlation. Where no redshift was available a 
temperature of $5\,{\rm keV}$ was assumed.\\ 
({\bf 9}) Energy flux from the cluster, with the absorption by galactic
interstellar absorption removed. The units are 
$10^{-12}\,{\rm erg}\,{\rm cm}^{-2}\,{\rm s}^{-1}$ 
in the $0.1-2.4\,{\rm keV}$ band.\\ 
({\bf 10}) X-ray luminosity of the cluster in its rest-frame, derived 
using the procedure described in section 4. The units are 
$10^{44}\,{\rm erg}\,{\rm s}^{-1}$ in the energy band $0.1-2.4\,{\rm keV}$. 
We assume ${\rm H}_{o}=50\,{\rm km}\,{\rm s}^{-1}\,{\rm Mpc}^{-1}$ 
and ${\rm q}_{o}=0.5$.\\
({\bf 11}) Optical identification of the cluster. The meaning of the prefix to 
the identification and the appropriate reference are given in Table 4. For five 
sources identified as double clusters, the label 'dc' has been appended to the
identification.\\
({\bf 12}) Redshift of the cluster.\\
({\bf 13}) Label for the reference from which the redshift was obtained:\\
{\bf 1.} Romer (1995) {\bf 2.} H. B\"{o}hringer and L. Guzzo 1999, ESO Key 
Project, private communication) {\bf 3.} Struble \&  Rood (1999) 
{\bf 4.} Katgert et al. (1996) {\bf 5.} Quintana \& Ramirez (1995) 
{\bf 6.} Fetisova et al. (1993) {\bf 7.} De Grandi et al. (1999) {\bf 8.} 
H. Andernach 1989,"A Compilation of Measured Redshifts of Abell Clusters", 
unpublished, private communication {\bf 9.} Crawford et al. (1995) 
{\bf 10.} Huchra et al. (1999)
{\bf 11.} Shectman (1985) {\bf 12.} Mazure et al. (1996) {\bf 13.} 
Collins et al.
 (1995) {\bf 14.} Dalton et al. (1997) {\bf 15.} De Vaucouleurs et al. (1991) 
 {\bf 16.} Stocke et al. (1991) {\bf 17.} Postman et al. (1992) {\bf 18.} 
 Abell, Corwin, \&  Olowin (1989) {\bf 19.} Katgert et al. (1998) 
 {\bf 20.} Ebeling \&  Maddox (1995) {\bf 21.} Muriel, Nicotra, \&  Lambas 
 (1995) {\bf 22.} Shectman et al. (1996)
{\bf 23.} Lauberts \&  Valentijn (1989) {\bf 24.} Da Costa et al. (1998) 
{\bf 25.} Dalton, Efstathiou, \& Sutherland (1994) {\bf 26.} Di Nella et al. 
(1996) {\bf 27.} Ratcliffe et al. (1998) {\bf 28.} Given in NED, reference 
uncertain.\\ 
({\bf 14}) Number of galaxy redshifts used in deriving the cluster redshift.\\

\begin{deluxetable}{ccc}
\tabletypesize{\small}
\tablenum{4}
\tablecolumns{3}
\tablecaption{Explanation with appropriate references of the prefixes to
 the source optical identification, which is given in column 11 of Table 3.} 
\tablehead{  
\colhead{Prefix} & \colhead{Source of Optical Identification} & 
\colhead{Reference} \\
}
\startdata 
  A & Abell catalog & Abell et al. (1989) \\
  S & Abell catalog: southern supplementary clusters & Abell et al. (1989) \\
  AM & Cambridge APM catalog & NED: Arp \&  Madore (1987) \\
  APMCC & Cambridge APM cluster catalog & Dalton et al. (1997) \\
  CID & EFAR survey & Barton et al. (1996) \\
  EDCC & Edinburgh-Durham cluster catalog & Lumsden et al. (1992) \\
  EMSS & EINSTEIN medium sensitivity survey & Stocke et al. (1991) \\
  ESO &  European Southern Observatory & No reference in NED   \\
  HCG &  Hickson compact group catalog & Hickson (1982), Barton et al. (1996) \\
  SCG & Catalog of COSMOS southern compact groups & Prandoni, Iovino \& 
  MacGillivray (1994) \\
  NGC & Cluster dominated by NGC galaxy & NED \\
  SH & Cluster catalog from Shane-Wirtanen galaxy counts & Shectman (1985) \\
\enddata  
\end{deluxetable}  

\begin{figure*}
\centerline{\psfig{figure=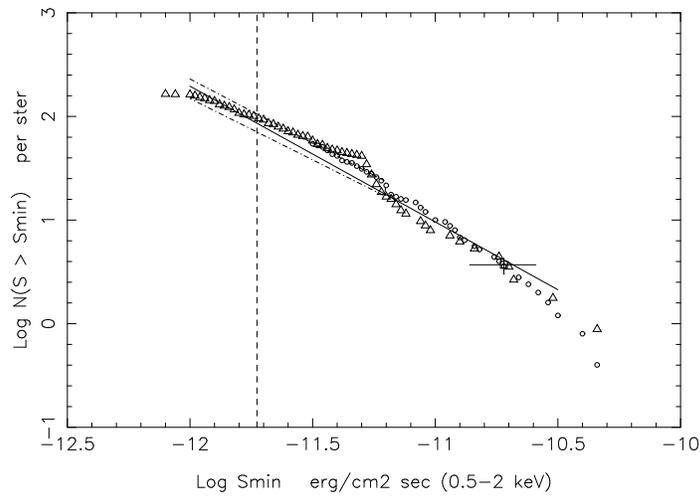,width=4.0in,angle=270}}
\caption{The log N-log S distribution of the SGP cluster sample is
traced by the symbols $\triangle $. The distribution obtained by 
De Grandi et al. (1999) is shown by the symbols $\circ $. The solid line is the
fit of Ebeling et al. (1998) and the dot-dash line shows limits set by Rosati 
et al. (1995) using the EMSS survey results. The point ($\Box $) with error 
bars was obtained from Piccinotti et al. (1982). For comparison purposes all
fluxes in this plot refer to the energy band $0.5-2.0\,{\rm keV}$. The vertical
dashed line corresponds to a flux limit in the $0.1-2.4\,{\rm keV}$ band
of $3\times 10^{-12}\,{\rm erg}\,{\rm cm}^{-2}\,{\rm s}^{-1}$.} 
\label{f12}
\end{figure*}

Images of these clusters, showing overlays of the hard band X-ray contours upon optical images reconstructed from the COSMOS 
digital data base, are accessible over the Internet at 
(http://wave.xray.mpe.mpg.de/publications/papers/2001/RASS-SGP-clusters/
xray-optical-images). 
They comprise 134 Abell clusters, 15 clusters 
found in other catalogs, and 37 newly discovered clusters. The number of Abell
clusters in the SGP survey area is 1215, including 332 southern supplementary 
clusters, Therefore the Abell clusters in the SGP catalog (Table 3) comprise 
11.0 percent of all Abell clusters in the SGP survey area. The Abell
fraction of the SGP cluster sample (72\%) is in good agreement with that of the 
BCS (Ebeling et al. 1999, 2000). 
We have compared the catalog in Table 3 with the REFLEX catalog of bright
clusters published by De Grandi et al. (1999), which contains 54 clusters lying
in the SGP region. Applying their energy flux threshold of 
$5-6.6\times 10^{-12}\,{\rm erg}\,{\rm cm}^{2}\,{\rm s}^{-1}$ in the 
$0.1-2.4\,{\rm keV}$ band to Table 3, we
find all but one cluster from the list of De Grandi et al. (1999). The exception
is A3866, which we have rejected tentatively as a source whose X-ray flux may 
be dominated by an
AGN (section 3.5.2). However Table 3 contains 20 clusters above this threshold,
which are not in the list of De Grandi et al. (1999). The reason is that these 
sources were not in the RASS-1 source list, but were found by the RASS-2 
analysis as a consequence of the improvements, in particular lowering source 
detection thresholds and searching for sources in the hard energy band, which
are described in section 3.1 and by Voges et al. (1999). These 20 sources
are NGC 720, NGC 1132, five newly discovered clusters, SH518, and the Abell
clusters 194, 2410, 2442, 2474, 2496, 2554, 2556, 2566, 3984, 4010, 4068 and 
S1136. The SGP catalog contains 113 clusters having fluxes below the threshold 
of the De Grandi et al. (1999) study. 

The log N-log S distribution for the SGP cluster sample is shown in
Figure \ref{f12}, where it is compared with the results of other surveys.
To avoid any errors caused by converting hard band ($0.5-2.0\,{\rm keV}$) 
fluxes in 
these surveys to broad band ($0.1-2.4\,{\rm keV}$) fluxes, we have made this
comparison in the hard band. The circular symbols ($\circ $) are the
result of the survey of clusters by De Grandi et al. (1999), who reanalysed 
clusters detected in the RASS-1 data base over a $2.5\,{\rm ster}$ area of the 
southern sky and made first use of redshifts obtained from the ESO Key Project.
Figure \ref{f12} includes also one point from the HEAO-1 A-2 cluster 
survey (Piccinotti et al. 1982), the fit to the log N-log S curve obtained by 
Ebeling et al. (1998) from the ROSAT Brightest Cluster sample (BCS), and a box 
setting limits at low fluxes, derived from an analysis of EMSS results 
(Rosati et al. 1995). There are significant differences among these results
at energy fluxes less than 
$1\times 10^{-11}\,{\rm erg}\,{\rm cm}^{-2}\,{\rm s}^{-1}$. There is a
flattening of the SGP log N-log S distribution at energy fluxes less than about
$1.8\times 10^{-12}\,{\rm erg}\,{\rm cm}^{-2}\,{\rm s}^{-1}$, equivalent to
a flux of $3\times 10^{-12}\,{\rm erg}\,{\rm cm}^{-2}\,{\rm s}^{-1}$ in the 
ROSAT broad band ($0.1-2.4\,{\rm keV}$), which is the completeness limit 
derived in section 6. Applying this limit we obtain a sample of 112 clusters, 
of which 110 have measured redshifts. The sources of the incompleteness of the 
SGP catalog at low fluxes have been discussed in section 3.5.5. 

In addition there is a steepening of the SGP log N-log S curve
over a short range of hard band flux near 
$5\times 10^{-12}\,{\rm erg}\,{\rm cm}^{-2}\,{\rm s}^{-1}$, a characteristic 
evident to a lesser extent also in the results of De Grandi et al. (1999). This
results in the distribution being higher than the fit of Ebeling et al. (1998) 
and being at the upper limit set by Rosati et al. (1995), within the range 
$3-5\times 10^{-12}\,{\rm erg}\,{\rm cm}^{-2}\,{\rm s}^{-1}$. This may be a
statistical fluctuation, but as we show in section 7, there is 
significant large-scale structure in the SGP field out to a redshift of
${\rm z}\sim 0.15$, which may be one cause of this feature in the SGP 
log N-log S distribution (Figure \ref{f12}).

\begin{figure*}
\centerline{\psfig{figure=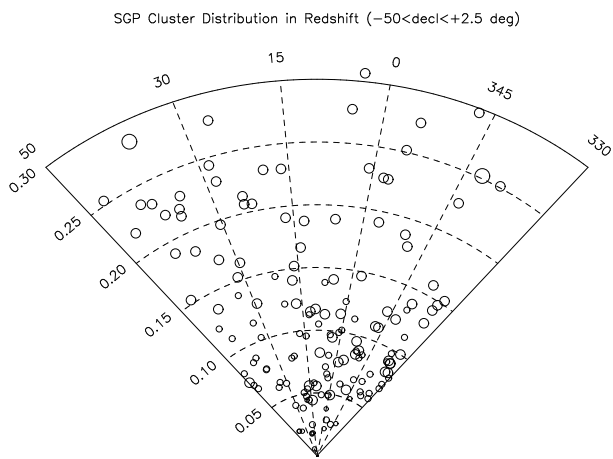,width=4.0in,angle=270}}
\caption{A 'wedge' diagram showing the distribution in redshift of all the
clusters in the catalog (Table 3), for which measured redshifts
are available. The symbol size is a measure of X-ray luminosity in the hard
energy band.}
\label{f13}
\end{figure*}

The optical observing
program and a search of the literature yielded redshifts for 157 of the 186
clusters
in Table 3. The distribution of the clusters in redshift space is summarised in
the 'wedge' diagram shown in Figure \ref{f13}, in which the clusters 
are projected onto a plane whose co-ordinates are cluster redshift and right 
ascension. In this figure the symbol size is a measure of the cluster X-ray
luminosity in the $0.1-2.4\,{\rm keV}$ band.  

Table 3 contains five sources, labelled 'dc' in column 11, for which the
X-ray and optical data provide compelling evidence that they are interacting
double clusters. In contrast the pair of objects RXCJ0108.8-1524 and 
RXCJ0108.9-1537, which in projection are close, do not constitute a double 
cluster as they are well separated in redshift space. The former is A151, which 
has a redshift of 0.0537 (H. Andernach 1989, unpublished, private 
communication), whereas spectroscopic observation of the latter
yield two redshifts which tie the galaxies to A151 but nine having a mean 
of $0.0981\pm 0.0027\,({\rm rms})$ (Katgert et al. 1996, 1998). 
 
\section*{6. TESTS OF SAMPLE COMPLETENESS}
In section 3.5.5 it was estimated from a comparison of the number of clusters 
detected in the SGP with the CSEARCH predictions, that the sample completeness
was 0.79, and it was concluded that the missing ones were mainly distant 
clusters of small angular extent. The effect is to flatten the log N-log S
distribution at low fluxes (Figure \ref{f12}), and using the following  
arguments we have established that the sample is essentially complete at fluxes
greater than $3.0\times 10^{-12}\,{\rm erg}\,{\rm cm}^{-2}\,{\rm s}^{-1}$. 

First, we compare our log N-log S distribution with the results of deeper 
surveys, and of the wider field REFLEX and BCS surveys. At
$3.0\times 10^{-12}\,{\rm erg}\,{\rm cm}^{-2}\,{\rm s}^{-1}$ in the SGP
distribution (Figure \ref{f12}) we obtain ${\rm N}=100\pm 9\,{\rm ster}^{-1}$. 
For deeper surveys, which have smaller fields and therefore limited statistics
at this flux, we obtain ${\rm N}=87\pm 15\,{\rm ster}^{-1}$ 
(Rosati et al. 1998),
${\rm N}=70\pm 10\,{\rm ster}^{-1}$ (Vikhlinen et al. 1998), 
${\rm N}=125^{+98}_{-66}\,{\rm ster}^{-1}$ (Henry et al. 2001), and
${\rm N}=87^{+69}_{-38}\,{\rm ster}^{-1}$ (Gioia et al. 2001). For the REFLEX
survey we obtain ${\rm N}=94\pm 5\,{\rm ster}^{-1}$ (B\"{o}hringer et al. 2001)
and for the BCS survey ${\rm N}=95\pm 5\,{\rm ster}^{-1}$ (Ebeling et al. 1998).
We find no evidence that the SGP result is inconsistent with these surveys 
at this flux threshold.

\begin{figure*}
\centerline{\psfig{figure=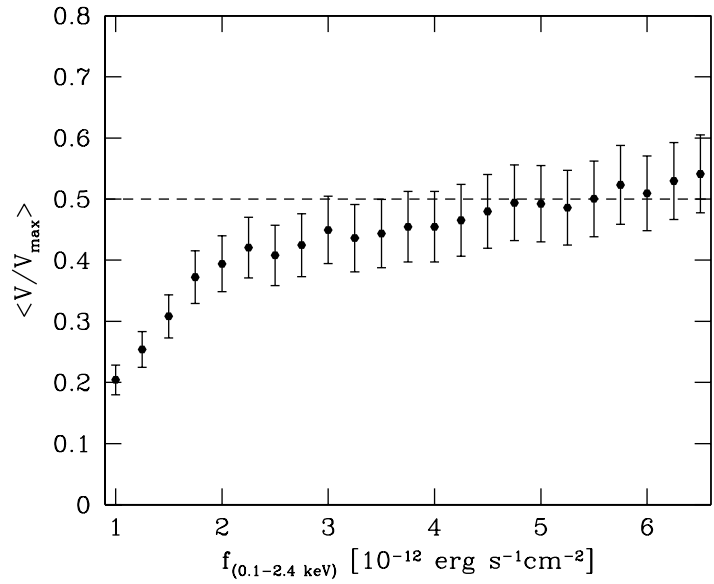,width=4.0in,angle=0}}
\caption{Mean ${\rm V}/{\rm V}_{\rm max}$ values as a function of flux limit. 
The error bars are the formal $2\sigma $ Poisson errors. The dashed line
corresponds to the ideal case of a complete sample. It is seen that
below $3.0\times 10^{-12}\,{\rm erg}\,{\rm cm}^{-2}\,{\rm s}^{-1}$ the
sample becomes incomplete.}
\label{f14}
\end{figure*}

Second, we apply a standard measure of the formal completeness of a sample,
namely the ${\rm V}/{\rm V}_{\rm max}$ test (e.g., Schmidt 1968, Avni \&  
Bahcall 1980). For the computation of the maximum volume ${\rm V}_{max}$ the 
effective area, the local
variation of the X-ray flux limit (both for a minimum of 10 source
counts), and the ${\rm K}$ correction (see section 4) are taken into
account. The test is performed for subsamples with different X-ray
flux limits. The mean of the ${\rm V}/{\rm V}_{\rm max}$ values and their
$2\sigma$ standard deviations are plotted in Figure \ref{f14}. 
Note that the errors do not include cosmic variance. Below 
$2\times 10^{-12}\,{\rm erg}\,{\rm cm}^{-2}\,{\rm s}^{-1}$ the
deviations from the ideal value of 0.5 are significant, indicating
that the sample gets definitively incomplete below this flux
limit. Between $2\times 10^{-12}$ and $3.0\times 10^{-12}\,{\rm
erg}\,{\rm cm}^{-2}\,{\rm s}^{-1}$ it is difficult to make a definite
statement about significant deviations from the 0.5 line because the
large scale inhomogeneities in the spatial distribution of the
clusters (see section 7) start to modulate the averaged 
${\rm V}/{\rm V}_{\rm max}$
values. Note also that the measurements at different flux limits are
not statistically independent. A good estimate of the flux
completeness limit is $f_x^{\rm lim}=3\times 10^{-12}\,{\rm erg}\,{\rm
cm}^{-2}\,{\rm s}^{-1}$, because for fluxes brighter than this limit
the averaged ${\rm V}/{\rm V}_{\rm max}$ values are all within the 90 percent
confidence range (note that cosmic variance increase the given formal
errors).

\begin{figure*}
\centerline{\psfig{figure=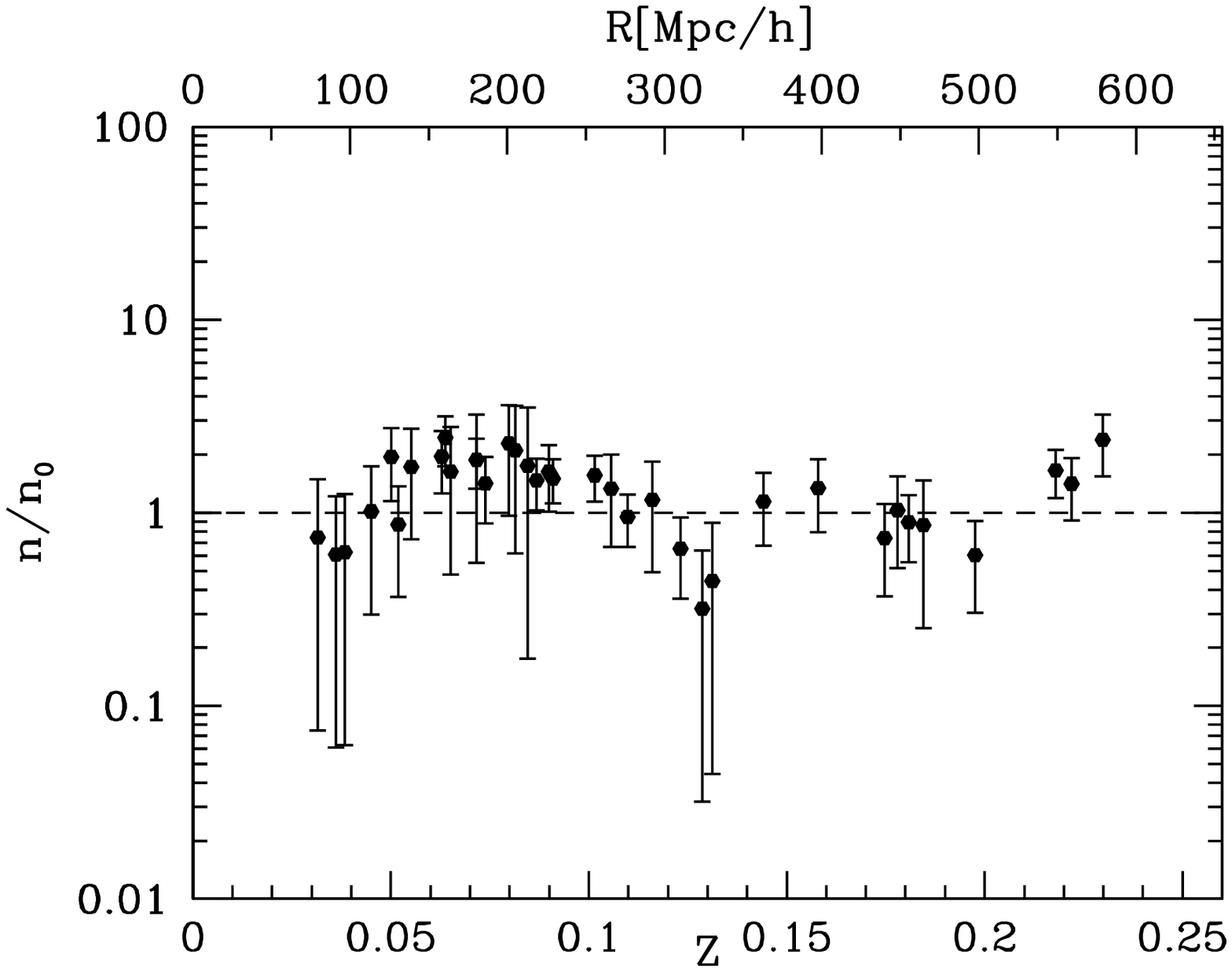,width=4.0in,angle=0}}
\caption{Comoving cluster number densities as a function of redshift
obtained from 10 subsamples covering the luminosity range from
$5.0\times 10^{43}$ to $2.0\times 10^{45}\,{\rm erg}\,{\rm
s}^{-1}$. The clusters have fluxes larger than $3.0\times
10^{-12}\,{\rm erg}\,{\rm cm}^{-2}\,{\rm s}^{-1}$ and at least 10
source counts. No indications for systematic large-scale density
gradients are seen, suggesting the absence of strong incompleteness
effects.}
\label{f15}
\end{figure*}

Third, we examine the behavior of the
radially averaged comoving cluster number density along the redshift
direction. If we assume that within the tested redshift range no large
evolutionary effects are present, one would expect an almost constant
cluster number density profile with superposed large-scale
inhomogeneities. Figure \ref{f15} shows the superposition of 
a set of cluster
number density profiles obtained from 10 volume-limited subsamples
within the luminosity range $5.0\times 10^{43}$ to $2.0\times
10^{45}\,{\rm erg}\,{\rm s}^{-1}$. Outside this luminosity interval
the sizes of the subsamples are smaller than 10 and are thus dominated
by small-number statistics. The subsamples are restricted to clusters
with at least 10 source counts and X-ray fluxes larger than $3.0\times
10^{-12}\,{\rm erg}\,{\rm cm}^{-2}\,{\rm s}^{-1}$, as obtained from the
${\rm V}/{\rm V}_{\rm max}$ test mentioned above. The effective survey area
covered by each subsample is taken into account. The individual
density profiles are normalized using the average cluster number
density estimated for each subsample. The bars in 
Figure \ref{f15} represent the formal $1\sigma$ Poisson errors, within 
which no systematic large-scale gradients of the
number density are found for ${\rm z}\le 0.25$.
 The comparatively large
high-density region found in the redshift interval $0.05-0.10$ is
composed mainly of several superclusters. A more detailed analysis of
the large-scale structure is given in section 7.

Finally, we have verified that the variation of RASS exposure time in the SGP
field (Figure \ref{f1}) has little effect on the effective area at a flux level
of $3.0\times 10^{-12}\,{\rm erg}\,{\rm cm}^{-2}\,{\rm s}^{-1}$. At the 
detection threshold of the GCA (10 source counts), this area is 98 percent of
the total area of the field.
\section*{7. LARGE-SCALE STRUCTURE IN THE SGP REGION}
\begin{figure*}
\centerline{\psfig{figure=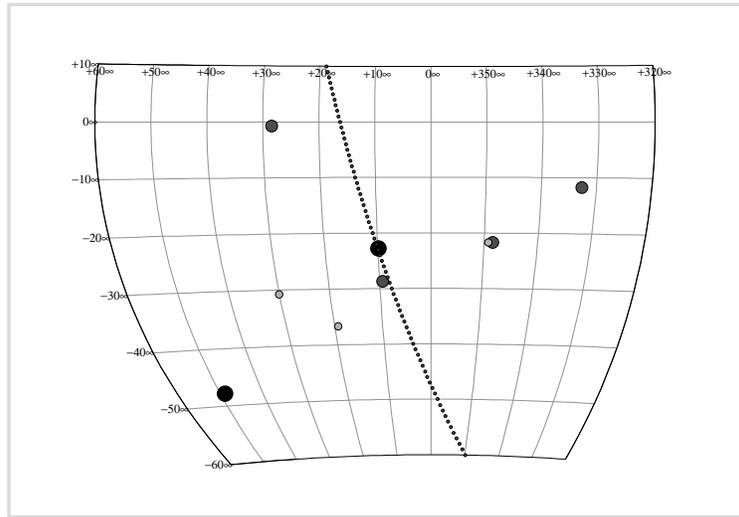,width=4.0in,angle=270}}
\caption{The map shows the locations of the centroids of superclusters (SC)
identified in Tully et al. (1992) in the SGP region. The symbol size is a rough
measure of distance, decreasing as redshift increases. The following notes
identify the superclusters and the centroid positions: Pisces-Cetus (10,-23),
Horologium-Reticulum (49,-48), Sculptor A (9,-29), Aquarius A (348,-22),
Aquarius B (349,-22), Aquarius-Capricornus (333,-12), Cetus (28,-1), 
Fornax (30,-31) and Sculptor B (19,-37). The dotted line traces the
supergalactic equator}.
\label{f16}
\end{figure*}

\begin{figure*}
\centerline{\psfig{figure=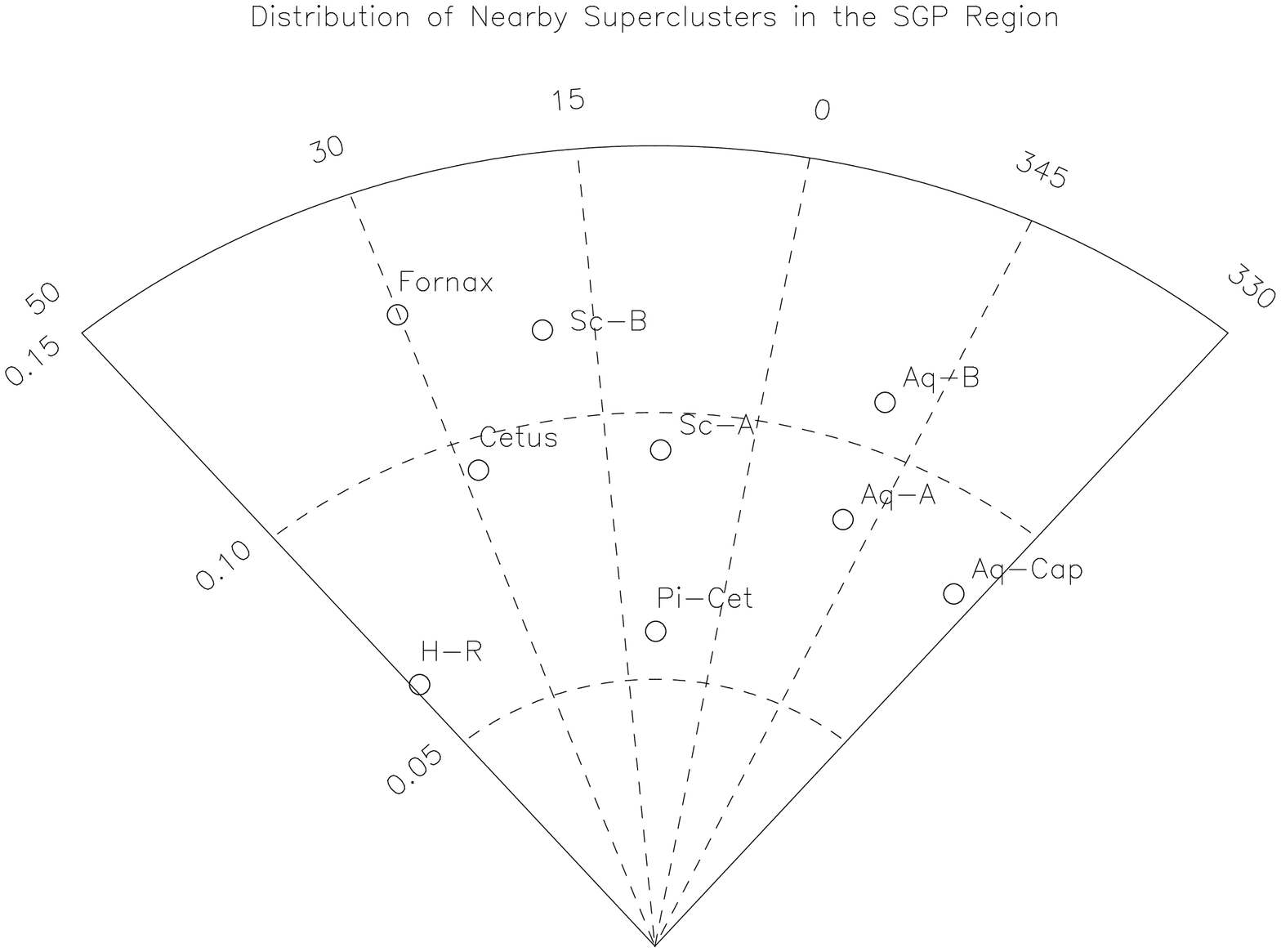,width=4.0in,angle=270}}
\caption{A 'wedge' diagram showing the distribution in redshift space of the
supercluster centroids shown in Figure \ref{f16}. The labels identify the
superclusters.}
\label{f17}
\end{figure*}

The 'wedge' diagram for the SGP cluster sample (Figure \ref{f13})
reveals that the spatial density of clusters at redshifts greater than 
$0.15$ is too low for the purpose of tracing large-scale structure, and we 
concentrate the discussion on clusters within this limit. A detailed 
examination of large-scale structure in the southern sky was made by Tully et 
al. (1992), who analysed the Abell,Corwin \& Olowin (1989) catalog of 
clusters. In this study measured redshifts were complemented by estimates based
on the magnitudes of the third and tenth brightest galaxies, having an 
accuracy of $\pm 22\% \,{\rm (rms})$ out to redshifts of about $0.2$ 
(Zamorani et al. 1992). We have condensed the results into a sky map of the 
centroids of superclusters identified in the SGP region 
(Figure \ref{f16}). Another 'wedge' diagram (Figure \ref{f17})
shows the distribution of these superclusters in redshift space.
 
\begin{figure*}
\centerline{\psfig{figure=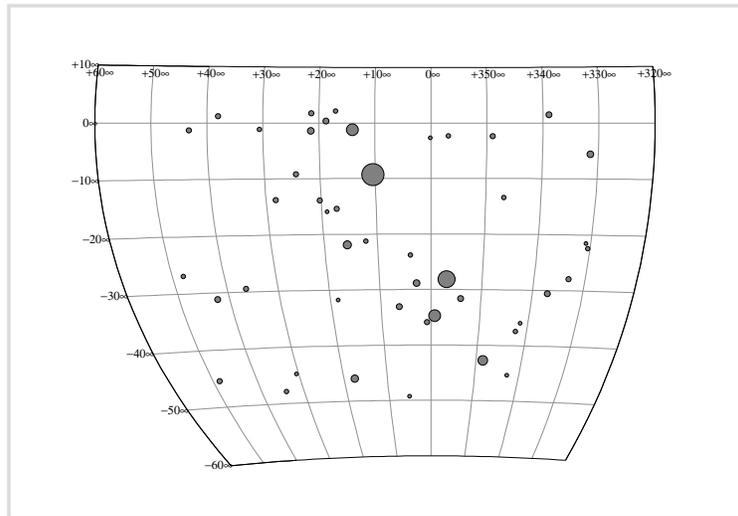,width=4.0in,angle=270}}
\caption{A sky map of clusters of galaxies detected by the ROSAT sky survey in a
region centered on the south galactic pole and having an area of 
$1.013\,{\rm ster}$.
The map shows only clusters having a redshift ${\rm z}<0.075$. The symbol size 
is a 
measure of the X-ray count-rate in the hard energy band ($0.5-2\,{\rm keV}$).}
\label{f18}
\end{figure*}

The dotted line in Figure \ref{f16} traces the supergalactic equator,
defined by the plane of a 'disk-like' structure, the Local Supercluster,
identified by De Vaucouleurs \& De Vaucouleurs (1964) using measurements of   
the radial velocities of a large sample of local galaxies. The maximum values of
these velocities were around $10000\,{\rm km}\,{\rm s}^{-1}$ 
(${\rm z}\simeq 0.03$). 
Tully et al. (1992) have presented persuasive evidence that the Pisces-Cetus 
supercluster is likewise a flattened structure extending along the 
supergalactic equator, which is an extension of the Local Supercluster out to 
redshifts of at least $0.06$. This idea is supported by our X-ray survey of 
clusters in the SGP. Figure \ref{f18} is a sky map of all clusters
in the SGP catalog (Table 3) having a redshift of less than $0.075$, which
shows clearly a concentration of clusters around the supergalactic equator 
(Figure \ref{f16}). Of the 36 clusters identified by Tully (1987) as 
members of the Pisces-Cetus supercluster, the following seven appear in Table 3
and have a redshift ${\rm z}<0.075$: A14, A85, A119, A133, A147, A151, and A168.
The six brightest clusters in Figure \ref{f18}, having hard-band
X-ray count rates greater then $0.9\,{\rm ct}\,{\rm s}^{-1}$, lie close to the 
supergalactic 
equator. Five of these clusters, A85, A119, A133, A4059 and S1101, have 
redshifts between $0.0444$ and $0.0564$, consistent with their being members of
the Pisces-Cetus supercluster. The fifth, A4038, is nearer at a redshift of 
$0.0283$, placing it at the edge of the Local Supercluster. The Pisces-Cetus 
supercluster is evident also in the 'wedge' diagram 
(Figure \ref{f13}), in which the group of X-ray clusters at 
${\rm R}.{\rm A}.\sim 15^{\rm o}$ and ${\rm z}\sim 0.05$ is close to the 
location of the supercluster
center.

\begin{figure*}
\centerline{\psfig{figure=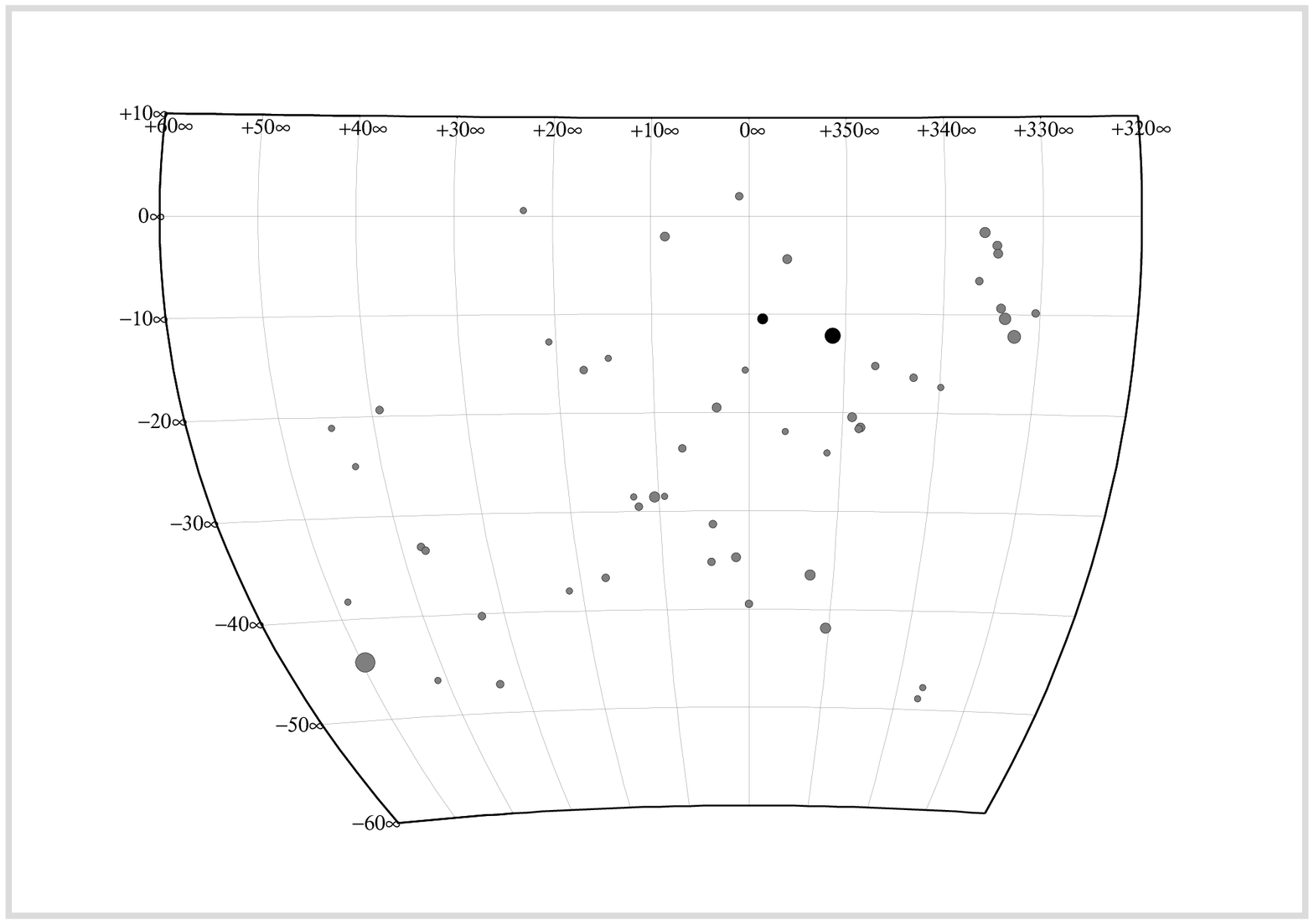,width=4.0in,angle=270}}
\caption{A sky map of those clusters in the ROSAT SGP sample, which have
redshifts in the range $0.075<{\rm z}<0.125$. The two black circles in the NW
quadrant are A2597 and A2670, identified by Tully (1987) as members of the
Aquarius supercluster association. }
\label{f19}
\end{figure*}

More distant structures are difficult to identify in the X-ray survey. Tully et
al. (1992) identified a structure orthogonal to the supergalactic equator,
which includes the superclusters Aquarius A and B, Aquarius-Capricornus, and
Sculptor A and B. The location of this structure, which extends from 
${\rm z}\sim 0.05$ to ${\rm z}\sim 0.12$ 
($\sim 130-330{\rm h}^{-1}\,{\rm Mpc}$), may be traced in
Figures \ref{f16} and \ref{f17}.  To search for evidence of
this structure in the X-ray data, we have constructed a sky map 
(Figure \ref{f19}) which includes only X-ray clusters in the range
$0.075<{\rm z}<0.125$. The map shows an elongated feature in the NW quadrant, 
which is
perpendicular to the supergalactic equator. The argument that this feature
traces the structure evident in the findings of of Tully et al. (1992) is
supported by two observations.  First, the group of X-ray clusters in the NW 
corner of Figure \ref{f19} belongs to the Aquarius-Capricornus 
supercluster. A correlation in sky position is evident upon comparing  
Figures \ref{f19} and \ref{f16}, and a correlation in 
redshift space is seen upon comparing Figures \ref{f13} and 
\ref{f17}. Second, the two bright clusters A2597 and A2670, 
highlighted as black circles in the NW quadrant, were identified by Tully (1987)
as members of the Aquarius supercluster. 

Batuski et al. (1999) have performed 
redshift measurements recently which establish that Aquarius A and B are 
essentially one filamentary structure almost aligned with the line-of-sight
between ${\rm z}=0.08$ and ${\rm z}=0.15$. This concentration of Abell clusters
on the sky
near ${\rm \alpha }=350^{\rm o}$, ${\rm \delta }=-22^{\rm o}$ 
(Batuski et al. 1999) is not evident in
the X-ray data (Figure \ref{f19}). However, care is needed in 
interpreting
this result as it may be biased by the relatively low exposure time in this
part of the RASS (Figure \ref{f1}).

Therefore we conclude that, at least for redshifts out to ${\rm z}\sim 0.15$, 
there is
significant large-scale structure in the SGP field. However, there is one part 
of this structure which is not strongly evident in the X-ray maps shown in
Figure \ref{f18} and Figure \ref{f19}. This is the 
Horologium-Reticulum supercluster (Tully et al. 1992), a rich association of
clusters having its centroid at ${\rm z}=0.063$ and located at the position 
shown in Figure \ref{f16}. There is only one bright X-ray cluster in our
survey, A3113 at ($49.49,-44.24$, Figure \ref{f19}) and ${\rm z}=0.0754$,
which may be associated with this supercluster. We cannot rule out the 
possibility that the supercluster is inconspicuous because of its location close
to the boundary of the SGP survey region.

\section*{\normalsize \bf 8. SUMMARY AND CONCLUSIONS}
A systematic search of the ROSAT all-sky survey for clusters of galaxies has
been performed in an area of $1.013\,{\rm ster}$ centered on the South Galactic
Pole.
The search was made by correlating the list of RASS sources detected in the hard
X-ray band ($0.5-2.0\,{\rm keV}$) with the projected surface density of 
galaxies 
derived from the COSMOS digitised UK Schmidt IIIa-J survey of the southern sky.
The analysis yielded a list of 3236 candidates, of which $\sim50\,{\rm percent}$
should be chance associations with non-cluster sources, mainly AGN and stars,
and which should be 90 percent complete in containing clusters. The study was 
bounded during X-ray analysis by applying a lower limit of 
$0.08\,{\rm ct}\,{\rm s}^{-1}$ to the count-rate in the hard band, yielding a 
list of 477 candidates.

Non-cluster sources were removed by applying tests using the source X-ray 
extent and hardness ratio, by scrutiny of COSMOS finding charts and overlays of
the RASS and COSMOS images, and by correlation with such catalogs as the NASA 
Extragalactic (NED), SIMBAD, TYCHO and Veron data bases. This was performed in 
as objective a manner as possible, requiring an iterative sequence of 
procedures.
The search resulted in 186 clusters, comprising 134 Abell clusters, 15 clusters 
found in other catalogs, and 37 newly discovered clusters. The catalog shown in 
Table 3, and the overlays of the RASS hard X-ray and COSMOS optical images, are 
accessible over the Internet 
(http://wave.xray.mpe.mpg.de/publications/papers/2001/RASS-SGP-clusters/xray-
optical-images). 
The catalog has a minimum energy flux of 
$\sim 1.5\times 10^{-12}\,{\rm erg}\,{\rm cm}^{-2}\,{\rm s}^{-1}$ in the band 
$0.1-2.4\,{\rm keV}$. Tests show that a sample of 112 clusters, obtained when 
a flux limit of $3.0\times 10^{-12}\,{\rm erg}\,{\rm cm}^{-2}\,{\rm s}^{-1}$ 
is applied, is essentially complete.  We estimate the completeness of the 
overall catalog at a flux limit of 
$1.5\times 10^{-12}\,{\rm erg}\,{\rm cm}^{-2}\,{\rm s}^{-1}$ to be 
$\leq 80\,{\rm  percent}$.

In support of studies of cluster evolution and large-scale structure using the
results of the survey, a series of optical observing campaigns at the AAT and 
at SAAO was conducted to obtain cluster redshifts. In addition we have used a
number of redshifts obtained in programs at ESO in support of the REFLEX 
project. Using these campaigns and an extensive search of the literature 
we have obtained redshifts 
for 157 of the clusters in the catalog (Table 3). Redshifts 
are available for 110 of the 112 clusters in the complete flux-limited sample.

Examination of the spatial distribution of the SGP X-ray cluster sample 
corroborates earlier studies of large-scale structure, which were based on
optical observations (De Vaucouleurs \&  De Vaucouleurs 1964, Tully et al. 
1992). The basic features are a flattened structure in the 
supergalactic plane, reaching out to ${\rm z}\sim 0.07$ and embracing the 
Pisces-Cetus supercluster, and an orthogonal structure at $0.07<{\rm z}<0.15$ 
which includes the Aquarius, Aquarius-Capricornus and Sculptor A and B 
superclusters.

\acknowledgements We thank Prof. Tr\"{u}mper and the ROSAT team for the
unstinting support, the myriad valuable discussions, the access to the RASS data
base, the use of the RASS and EXSAS software, and the use of the powerful
computing system under Joachim Paul and his team, which have made this work
possible. The ROSAT Project is supported by the Bildungsministerium f\"{u}r
Bildung und Forschung (BMBF/DLR) and the Max-Planck-Gesellschaft (MPG).
In equal measure we thank the Royal Observatory Edinburgh and the
Naval Research Laboratory (NRL) for the intensive effort that went into the 
creation of the
COSMOS digitised optical data base and its installation at MPE, which was an 
essential factor in identifying clusters in the RASS data base. Our gratitude
goes also to the Anglo-Australian Telescope, the South African Astronomical
Observatory and the European Southern Observatory for the expert assistance
provided to the optical observations crucial to this project. We thank the
REFLEX project for its cooperation in exchanging redshifts, and in particular
Luigi Guzzo of the Osservatorio Astronomica di Brera for his analysis of those 
ESO Key Project observations, which yielded 14 redshifts used in this study. 
Finally, this research has been supported by the Office of Naval Research and 
NRL, and by the National Aeronautics and Space Administration
through an Astrophysics Data Program grant (S57759-F).

\newpage

\begin{deluxetable}{cccccccccccccc}                                     
\tabletypesize{\tiny}                                                   
\tablewidth{0pc}                                                        
\setlength{\tabcolsep}{0.02in}                                          
\tablenum{3}                                                            
\tablecolumns{14}                                                       
\tablecaption{Catalog of clusters in the SGP}                           
\tablehead{                                                             
\colhead{Source} & \colhead{RA  (J2000)} & \colhead{Decl  (J2000)} &    
\colhead{Count  rate} & \colhead{HR} & \colhead{Time} &                 
\colhead{$N_{H}$} & \colhead{$T_{x}$} & \colhead{$f_{x}$} &             
\colhead{$L_{x}$} & \colhead{Identification} & \colhead{Redshift} &     
\colhead{Ref(z)} & \colhead{$N_{g}$}\\                                  
\colhead{(1)} & \colhead{(2)}	 & \colhead{(3)} & \colhead{(4)} &        
\colhead{(5)} & \colhead{(6)} & \colhead{(7)} & \colhead{(8)} &         
\colhead{(9)} & \colhead{(10)} & \colhead{(11)} & \colhead{(12)} &      
\colhead{(13)} & \colhead{(14)}                                         
}                                                                       
\startdata                                                              
 RXCJ2201.8-2226 & 22 01 50.5 & -22 26 37  & 0.286$\pm $0.043 & 0.82$\pm $0.30 & 237 & 2.61 &  3.554$^{e}$ &   5.75 &  1.193 & S987                 & 0.0693 &  8    &  20 \\       
 RXCJ2202.1-0949 & 22 02 07.2 & -09 49 48  & 0.280$\pm $0.038 & 1.15$\pm $0.31 & 307 & 4.20 &  4.039$^{e}$ &   5.89 &  1.629 & A2410                & 0.0801 & 17    &   1 \\       
 RXCJ2203.8-2130 & 22 03 50.6 & -21 30 31  & 0.139$\pm $0.030 & 0.11$\pm $0.19 & 227 & 2.45 &  2.809$^{e}$ &   2.77 &  0.644 &                      & 0.0732 &  1    &   3 \\       
 RXCJ2205.6-0535 & 22 05 39.6 & -05 35 29  & 0.674$\pm $0.109 & 0.79$\pm $0.12 & 191 & 4.68 &  3.800$^{ }$ &  14.40 &  2.103 & A2415                & 0.0582 &  2    &   3 \\       
 RXCJ2210.3-1210 & 22 10 20.0 & -12 10 33  & 0.746$\pm $0.056 & 1.04$\pm $0.08 & 278 & 3.92 &  6.000$^{ }$ &  15.70 &  4.642 & A2420                & 0.0831 &  1    &   3 \\       
 RXCJ2213.0-2753 & 22 13 05.0 & -27 53 56  & 0.425$\pm $0.034 & 0.14$\pm $0.17 & 205 & 1.40 &  3.760$^{e}$ &   8.28 &  1.373 &                      & 0.0620 &  1    &   2 \\       
 RXCJ2214.4-1701 & 22 14 26.3 & -17 01 24  & 0.198$\pm $0.020 & 0.79$\pm $0.24 & 305 & 2.61 &  5.845$^{e}$ &   4.02 &  4.016 & A3847                & 0.1530 &  3    &   1 \\       
 RXCJ2214.5-1022 & 22 14 32.4 & -10 22 12  & 0.594$\pm $0.052 & 0.95$\pm $0.09 & 280 & 3.84 &  6.578$^{e}$ &  12.50 &  5.343 & A2426                & 0.1001 &  1    &   4 \\       
 RXCJ2216.2-0920 & 22 16 15.2 & -09 20 18  & 0.405$\pm $0.041 & 0.87$\pm $0.09 & 274 & 5.24 &  4.869$^{e}$ &   8.78 &  2.567 & A2428                & 0.0825 &  1    &   2 \\       
 RXCJ2216.9-1725 & 22 16 56.4 & -17 25 30  & 0.425$\pm $0.034 & 0.35$\pm $0.12 & 277 & 2.28 &  6.969$^{e}$ &   8.56 &  6.163 &                      & 0.1301 &  1    &   2 \\       
 RXCJ2217.7-3543 & 22 17 43.7 & -35 43 32  & 0.329$\pm $0.035 & 0.32$\pm $0.11 & 319 & 1.10 &  6.865$^{e}$ &   6.42 &  5.929 & A3854                & 0.1474 & 13    &   9 \\       
 RXCJ2218.2-0350 & 22 18 17.1 & -03 50 01  & 0.427$\pm $0.047 & 0.92$\pm $0.14 & 273 & 5.73 &  5.375$^{e}$ &   9.39 &  3.267 & MS2215.7-0404 dc     & 0.0901 &  1    &   3 \\       
 RXCJ2218.6-3853 & 22 18 40.3 & -38 53 48  & 0.425$\pm $0.034 & 0.27$\pm $0.09 & 314 & 1.33 &  7.246$^{e}$ &   8.35 &  6.746 & ESO 344-G 019        & 0.1379 &  28   &   ? \\       
 RXCJ2218.8-0258 & 22 18 49.0 & -02 58 07  & 0.425$\pm $0.034 & 0.90$\pm $0.19 & 268 & 5.84 &  5.350$^{e}$ &   9.37 &  3.246 &                      & 0.0899 &  2    &   7 \\       
 RXCJ2223.9-0137 & 22 23 56.7 & -01 37 44  & 0.454$\pm $0.049 & 1.00$\pm $0.12 & 224 & 5.34 &  9.000$^{ }$ &   9.92 &  3.509 & A2440                & 0.0912 &  8    &  24 \\       
 RXCJ2225.8-0636 & 22 25 50.8 & -06 36 09  & 0.239$\pm $0.034 & 0.68$\pm $0.15 & 251 & 5.14 &  4.196$^{e}$ &   5.16 &  1.787 & A2442                & 0.0897 &  2    &  12 \\       
 RXCJ2226.1-4745 & 22 26 10.6 & -47 45 01  & 0.091$\pm $0.020 & 0.31$\pm $0.23 & 329 & 1.44 &  3.289$^{e}$ &   1.77 &  0.978 & A3876                & 0.1130 & 14    &   ? \\       
 RXCJ2227.4-4852 & 22 27 28.3 & -48 52 54  & 0.126$\pm $0.022 & 0.24$\pm $0.17 & 367 & 1.65 &  3.321$^{e}$ &   2.47 &  1.004 & APMCC782             & 0.0970 & 14    &   3 \\       
 RXCJ2227.8-3034 & 22 27 52.9 & -30 34 11  & 0.522$\pm $0.043 & 0.05$\pm $0.06 & 308 & 1.09 &  3.800$^{ }$ &  10.10 &  1.415 & A3880                & 0.0570 & 13    &  11 \\       
 RXCJ2231.2-3802 & 22 31 15.1 & -38 02 39  & 0.088$\pm $0.022 & 0.88$\pm $0.44 & 233 & 1.20 &  5.000$^{e}$ &   1.71 &        &                      &        &       &     \\       
 RXCJ2234.5-3744 & 22 34 30.4 & -37 44 05  & 0.573$\pm $0.051 & 0.09$\pm $0.07 & 241 & 1.22 &  7.900$^{ }$ &  11.20 & 11.986 & A3888                & 0.1590 &  8    &  70 \\       
 RXCJ2234.8-3033 & 22 34 53.5 & -30 33 56  & 0.119$\pm $0.027 & 0.64$\pm $0.28 & 305 & 1.11 &  6.983$^{e}$ &   2.32 &  6.182 & A3889                & 0.2512 &  8    &  19 \\       
 RXCJ2235.6+0128 & 22 35 40.6 & +01 28 21  & 0.510$\pm $0.080 & 0.76$\pm $0.12 & 167 & 5.81 &  4.113$^{e}$ &  11.20 &  1.702 & A2457                & 0.0594 &  3    &  18 \\       
 RXCJ2239.2-1719 & 22 39 16.4 & -17 19 46  & 0.102$\pm $0.033 & 0.76$\pm $0.33 & 110 & 2.99 &  2.559$^{e}$ &   2.05 &  0.503 & A2462                & 0.0751 & 17    &   1 \\       
 RXCJ2241.9-4235 & 22 41 56.7 & -42 35 51  & 0.085$\pm $0.022 & 0.49$\pm $0.25 & 259 & 1.66 &  5.000$^{e}$ &   1.68 &        & A3902                &        &       &     \\       
 RXCJ2243.0-2010 & 22 43 05.1 & -20 10 34  & 0.246$\pm $0.098 & 0.71$\pm $0.20 & 185 & 2.60 &  5.892$^{e}$ &   4.99 &  4.085 & A2474                & 0.1385 &  1    &   1 \\       
 RXCJ2244.5-1744 & 22 44 34.9 & -17 44 21  & 0.128$\pm $0.035 & 0.38$\pm $0.29 & 151 & 2.79 &  4.361$^{e}$ &   2.60 &  1.963 & A2478                & 0.1325 &  1    &   2 \\       
 RXCJ2246.1-3600 & 22 46 11.8 & -36 00 22  & 0.100$\pm $0.022 & 0.40$\pm $0.25 & 260 & 1.14 &  2.290$^{e}$ &   1.90 &  0.374 &                      & 0.0673 &  1    &   2 \\       
 RXCJ2247.2+0204 & 22 47 15.2 & +02 04 35  & 0.168$\pm $0.064 & 0.93$\pm $0.18 & 151 & 5.84 &  5.000$^{e}$ &   3.70 &        &                      &        &       &     \\       
 RXCJ2248.7-4431 & 22 48 43.4 & -44 31 41  & 0.431$\pm $0.047 & 0.47$\pm $0.10 & 208 & 1.79 & 11.946$^{e}$ &   8.59 & 22.736 & S1063                & 0.2520 &  7    &   ? \\       
 RXCJ2249.2-3727 & 22 49 17.6 & -37 27 26  & 0.154$\pm $0.028 & 0.23$\pm $0.20 & 253 & 1.13 &  1.388$^{e}$ &   2.66 &  0.096 & S1065                & 0.0289 & 23    &   2 \\       
 RXCJ2250.8-4521 & 22 50 48.9 & -45 21 33  & 0.104$\pm $0.026 & 0.44$\pm $0.30 & 203 & 1.88 &  1.898$^{e}$ &   1.99 &  0.226 & S1067                & 0.0511 & 21,22 &   6 \\       
 RXCJ2251.0-1624 & 22 51 00.4 & -16 24 21  & 0.255$\pm $0.052 & 0.83$\pm $0.16 & 103 & 3.16 &  5.426$^{e}$ &   5.24 &  3.342 & A2496                & 0.1221 &  1    &   3 \\       
 RXCJ2253.5-3343 & 22 53 34.2 & -33 43 21  & 0.155$\pm $0.029 & 0.29$\pm $0.16 & 195 & 1.24 &  7.133$^{e}$ &   3.04 &  6.504 & A3934                & 0.2250 &  8    &   6 \\       
 RXCJ2305.5-4513 & 23 05 34.7 & -45 13 11  & 0.220$\pm $0.094 & 0.11$\pm $0.20 & 159 & 1.69 &  5.132$^{e}$ &   4.34 &  2.919 & A3970                & 0.1253 &  1    &   3 \\       
 RXCJ2306.8-1324 & 23 06 51.7 & -13 24 59  & 0.200$\pm $0.043 & 0.51$\pm $0.24 & 139 & 3.07 &  5.000$^{e}$ &   4.10 &  0.765 &                      & 0.0659 &  2    &   2 \\       
 RXCJ2307.2-1513 & 23 07 15.5 & -15 13 41  & 0.242$\pm $0.063 & 1.12$\pm $0.12 &  90 & 2.78 &  4.894$^{e}$ &   4.92 &  2.599 & A2533                & 0.1110 &  5    &   1 \\       
 RXCJ2308.3-0211 & 23 08 23.1 & -02 11 32  & 0.200$\pm $0.029 & 1.11$\pm $0.22 & 354 & 4.29 & 10.221$^{e}$ &   4.26 & 15.544 & A2537                & 0.2966 &  1    &   2 \\       
 RXCJ2312.2-2129 & 23 12 16.3 & -21 29 35  & 0.335$\pm $0.051 & 0.63$\pm $0.20 &  86 & 2.03 &  4.100$^{ }$ &   6.65 &  3.532 & A2554                & 0.1111 &  3    &  28 \\       
 RXCJ2313.0-2138 & 23 13 02.2 & -21 38 07  & 0.273$\pm $0.070 & 0.41$\pm $0.15 &  88 & 2.03 &  4.700$^{ }$ &   5.43 &  1.746 & A2556                & 0.0865 &  3    &   2 \\       
 RXCJ2313.9-4243 & 23 13 58.2 & -42 43 59  & 1.185$\pm $0.085 & 0.20$\pm $0.06 & 183 & 1.85 &  5.337$^{e}$ &  23.50 &  3.211 & S1101                & 0.0564 &  1    &   4 \\       
 RXCJ2315.7-3746 & 23 15 44.2 & -37 46 53  & 0.295$\pm $0.060 & 0.54$\pm $0.17 & 128 & 1.50 &  7.709$^{e}$ &   5.82 &  7.852 & AM 2312-380/A3984    & 0.1786 &  1    &   3 \\       
 RXCJ2315.7-0222 & 23 15 46.0 & -02 22 38  & 0.425$\pm $0.034 & 0.80$\pm $0.18 & 337 & 4.18 &  2.024$^{e}$ &   8.73 &  0.268 & NGC 7556             & 0.0267 &  1    &   2 \\       
 RXCJ2316.1-2027 & 23 16 08.0 & -20 27 19  & 0.373$\pm $0.068 & 0.48$\pm $0.15 &  95 & 2.07 &  4.589$^{e}$ &   7.43 &  2.222 & A2566                & 0.0834 &  1    &   2 \\       
 RXCJ2321.4-2312 & 23 21 25.9 & -23 12 31  & 0.164$\pm $0.042 & 0.42$\pm $0.21 &  98 & 1.96 &  5.100$^{ }$ &   3.26 &  4.883 & A2580                & 0.1870 & 16    &   1 \\       
 RXCJ2321.5-4154 & 23 21 33.0 & -41 54 04  & 0.440$\pm $0.059 & 0.40$\pm $0.12 & 143 & 2.15 &  5.178$^{e}$ &   8.80 &  2.983 & A3998                & 0.0889 &  1    &   2 \\       
 RXCJ2325.3-1207 & 23 25 20.1 & -12 07 40  & 1.016$\pm $0.059 & 0.44$\pm $0.05 & 322 & 2.50 &  9.100$^{ }$ &  20.60 &  6.365 & A2597                & 0.0852 &  8    &   3 \\       
 RXCJ2326.2-2406 & 23 26 16.0 & -24 06 13  & 0.105$\pm $0.023 & 0.73$\pm $0.26 & 185 & 1.82 &  2.888$^{e}$ &   2.06 &  0.692 & APMCC895             & 0.0880 & 14    &   ? \\       
 RXCJ2331.1-3630 & 23 31 11.3 & -36 30 22  & 0.562$\pm $0.124 & 0.54$\pm $0.17 &  51 & 1.44 &  6.005$^{e}$ &  11.00 &  4.279 & A4010                & 0.0954 &  8    &   8 \\       
 RXCJ2336.2-3135 & 23 36 14.9 & -31 35 55  & 0.448$\pm $0.057 & 0.16$\pm $0.19 & 125 & 1.18 &  3.712$^{e}$ &   8.67 &  1.320 & S1136                & 0.0594 & 26,27 &   2 \\       
 RXCJ2337.6+0016 & 23 37 40.9 & +00 16 34  & 0.169$\pm $0.033 & 0.55$\pm $0.15 & 378 & 3.85 &  8.948$^{e}$ &   3.56 & 11.264 & A2631                & 0.2753 &  6    &   1 \\       
 RXCJ2338.2-4944 & 23 38 12.7 & -49 44 43  & 0.094$\pm $0.024 & 0.75$\pm $0.28 & 175 & 1.89 &  5.000$^{e}$ &   1.87 &        &                      &        &       &     \\       
 RXCJ2341.2-0901 & 23 41 17.3 & -09 01 34  & 0.168$\pm $0.044 & 0.95$\pm $0.16 & 346 & 2.55 &  8.182$^{e}$ &   3.41 &  9.067 & A2645                & 0.2517 &  8    &   5 \\       
 RXCJ2344.2-0422 & 23 44 16.2 & -04 22 07  & 0.425$\pm $0.034 & 0.69$\pm $0.07 & 335 & 3.54 &  4.691$^{e}$ &   8.81 &  2.340 &                      & 0.0786 &  1    &   2 \\       
 RXCJ2344.4-2153 & 23 44 28.9 & -21 53 34  & 0.115$\pm $0.028 & 0.61$\pm $0.23 & 198 & 1.88 &  3.603$^{e}$ &   2.27 &  1.234 & A2655                & 0.1122 &  1    &   3 \\       
 RXCJ2346.7-1028 & 23 46 42.4 & -10 28 11  & 0.129$\pm $0.024 & 0.38$\pm $0.20 & 340 & 2.78 &  5.872$^{e}$ &   2.63 &  4.093 & A2661                & 0.1911 &  3    &   2 \\       
 RXCJ2347.4-0218 & 23 47 24.5 & -02 18 49  & 0.180$\pm $0.070 & 1.22$\pm $0.23 & 343 & 3.62 &  1.197$^{e}$ &   3.08 &  0.065 & HCG 97               & 0.0221 &  1    &   2 \\       
 RXCJ2347.7-2808 & 23 47 43.2 & -28 08 32  & 2.546$\pm $0.097 & 0.35$\pm $0.03 & 310 & 1.55 &  4.250$^{e}$ &  49.90 &  1.843 & A4038                & 0.0293 &  8    &  44 \\       
 RXCJ2348.2-2849 & 23 48 13.0 & -28 49 01  & 0.103$\pm $0.021 & 0.09$\pm $0.16 & 309 & 1.45 &  6.077$^{e}$ &   2.03 &  4.408 & A4041/S1151 dc       & 0.2260 & 18    &   ? \\       
 RXCJ2350.1-0339 & 23 50 09.1 & -03 39 10  & 0.125$\pm $0.026 & 1.00$\pm $0.23 & 340 & 3.65 &  4.733$^{e}$ &   2.60 &  2.399 & A2664                & 0.1466 &  6    &   2 \\       
 RXCJ2351.6-2605 & 23 51 40.5 & -26 05 09  & 0.468$\pm $0.047 & 0.48$\pm $0.07 & 308 & 1.66 & 11.318$^{e}$ &   9.29 & 19.901 & A2667                & 0.2264 &  1    &   1 \\       
 RXCJ2354.2-1024 & 23 54 13.5 & -10 24 53  & 0.473$\pm $0.057 & 0.68$\pm $0.08 & 310 & 2.92 &  3.900$^{ }$ &   9.62 &  2.403 & A2670                & 0.0761 &  8    & 220 \\       
 RXCJ2355.1-2834 & 23 55 08.6 & -28 34 27  & 0.082$\pm $0.020 & -.09$\pm $0.16 & 300 & 1.49 &  5.000$^{e}$ &   1.61 &        & A4054                &        &       &     \\       
 RXCJ2356.0-0129 & 23 56 02.4 & -01 29 43  & 0.160$\pm $0.028 & -.01$\pm $0.13 & 367 & 3.58 &  5.000$^{e}$ &   3.32 &        &                      &        &       &     \\       
 RXCJ2357.0-3445 & 23 57 02.4 & -34 45 45  & 1.688$\pm $0.093 & 0.30$\pm $0.05 & 255 & 1.10 &  3.500$^{ }$ &  32.60 &  3.404 & A4059                & 0.0492 &  1    &   2 \\       
 RXCJ2359.9-3928 & 23 59 56.8 & -39 28 53  & 0.258$\pm $0.047 & 0.56$\pm $0.20 & 152 & 1.33 &  4.595$^{e}$ &   5.03 &  2.229 & A4068                & 0.1016 &  8    &   2 \\       
 RXCJ0000.4-0237 & 00 00 24.7 & -02 37 30  & 0.092$\pm $0.019 & 1.00$\pm $0.37 & 372 & 3.56 &  1.460$^{e}$ &   1.77 &  0.110 &                      & 0.0379 & 24    &   1 \\       
 RXCJ0001.6-1540 & 00 01 39.0 & -15 40 52  & 0.116$\pm $0.021 & 1.00$\pm $0.27 & 337 & 2.24 &  3.960$^{e}$ &   2.32 &  1.552 &                      & 0.1246 &  1    &   3 \\       
 RXCJ0003.1-3555 & 00 03 11.1 & -35 55 21  & 0.400$\pm $0.071 & 0.29$\pm $0.14 & 142 & 1.11 &  3.046$^{e}$ &   7.70 &  0.799 & A2717                & 0.0490 &  3    &  56 \\       
 RXCJ0003.2-0605 & 00 03 12.0 & -06 05 12  & 0.218$\pm $0.027 & 0.71$\pm $0.14 & 350 & 3.16 &  8.577$^{e}$ &   4.50 & 10.167 & A2697                & 0.2320 &  2    &   ? \\       
 RXCJ0003.8+0203 & 00 03 50.9 & +02 03 45  & 0.205$\pm $0.039 & 0.83$\pm $0.12 & 362 & 3.00 &  4.185$^{e}$ &   4.18 &  1.776 & A2700                & 0.0994 &  1    &   2 \\       
 RXCJ0006.0-3443 & 00 06 04.8 & -34 43 26  & 0.301$\pm $0.041 & 0.23$\pm $0.13 & 226 & 1.16 &  5.381$^{e}$ &   5.86 &  3.276 & A2721                & 0.1143 &  8    &  72 \\       
 RXCJ0007.4-2809 & 00 07 25.3 & -28 09 21  & 0.103$\pm $0.022 & 0.43$\pm $0.27 & 313 & 1.77 &  5.000$^{e}$ &   2.04 &        & A2726                &        &       &     \\       
 RXCJ0011.3-2851 & 00 11 19.5 & -28 51 41  & 0.612$\pm $0.055 & 0.39$\pm $0.09 & 254 & 1.84 &  4.426$^{e}$ &  12.10 &  2.034 & A2734                & 0.0625 &  3    &  80 \\       
 RXCJ0012.9-0853 & 00 12 54.7 & -08 53 03  & 0.090$\pm $0.020 & 0.32$\pm $0.17 & 336 & 3.32 &  5.000$^{e}$ &   1.86 &        &                      &        &       &     \\       
 RXCJ0013.6-1930 & 00 13 38.2 & -19 30 08  & 0.306$\pm $0.035 & 0.48$\pm $0.11 & 313 & 2.04 &  4.699$^{e}$ &   6.09 &  2.355 & A13                  & 0.0949 &  1    &   4 \\       
 RXCJ0014.3-3022 & 00 14 20.1 & -30 22 33  & 0.250$\pm $0.047 & 0.23$\pm $0.15 & 122 & 1.65 & 11.184$^{e}$ &   4.96 & 19.331 & A2744                & 0.3067 &  8    &  34 \\       
 RXCJ0015.5-2346 & 00 15 31.8 & -23 46 13  & 0.175$\pm $0.032 & 0.36$\pm $0.19 & 277 & 2.45 &  2.814$^{e}$ &   3.48 &  0.647 & A14                  & 0.0655 &  3    &   9 \\       
 RXCJ0016.3-3121 & 00 16 19.8 & -31 21 55  & 0.178$\pm $0.032 & 0.00$\pm $0.14 & 296 & 1.88 &  3.292$^{e}$ &   3.50 &  0.981 & A2751                & 0.0805 & 20    &   1 \\       
 RXCJ0017.5-3511 & 00 17 34.9 & -35 11 09  & 0.166$\pm $0.033 & 0.28$\pm $0.17 & 354 & 1.31 &  3.686$^{e}$ &   3.22 &  1.304 & A2755                & 0.0969 &  8    &  18 \\       
 RXCJ0019.0-2026 & 00 19 03.9 & -20 26 17  & 0.147$\pm $0.027 & 0.42$\pm $0.17 & 322 & 1.78 &  8.314$^{e}$ &   2.93 &  9.432 & S26                  & 0.2773 &  1    &   1 \\       
 RXCJ0020.5-4913 & 00 20 34.1 & -49 13 40  & 0.124$\pm $0.030 & 0.77$\pm $0.25 & 151 & 2.12 &  2.620$^{e}$ &   2.44 &  0.536 & A2764                & 0.0711 &  3    &  19 \\       
 RXCJ0020.7-2542 & 00 20 43.3 & -25 42 53  & 0.289$\pm $0.035 & 0.48$\pm $0.12 & 281 & 2.26 &  6.360$^{e}$ &   5.81 &  4.921 & A22                  & 0.1410 &  3    &   3 \\       
 RXCJ0024.0-1704 & 00 24 03.6 & -17 04 32  & 0.088$\pm $0.019 & 0.63$\pm $0.24 & 314 & 2.00 &  4.956$^{e}$ &   1.75 &  2.680 & A2768                & 0.1890 &  1    &   2 \\       
 RXCJ0025.5-3302 & 00 25 32.2 & -33 02 53  & 0.460$\pm $0.045 & 0.37$\pm $0.09 & 328 & 1.69 &  3.215$^{e}$ &   9.00 &  0.921 & S41                  & 0.0487 &  8    &   2 \\       
 RXCJ0028.6-2338 & 00 28 39.5 & -23 38 27  & 0.245$\pm $0.033 & 0.55$\pm $0.13 & 294 & 1.82 &  4.933$^{e}$ &   4.85 &  2.650 & A42                  & 0.1129 &  8    &   5 \\       
 RXCJ0030.6-2410 & 00 30 37.6 & -24 10 53  & 0.094$\pm $0.019 & 0.60$\pm $0.22 & 314 & 1.51 &  3.927$^{e}$ &   1.84 &  1.516 & A47                  & 0.1382 &  3    &   1 \\       
 RXCJ0033.8-0750 & 00 33 51.6 & -07 50 29  & 0.087$\pm $0.021 & 0.56$\pm $0.24 & 270 & 3.50 &  5.000$^{e}$ &   1.80 &        & A56                  &        &       &     \\       
 RXCJ0034.2-0204 & 00 34 13.9 & -02 04 31  & 0.412$\pm $0.033 & 0.67$\pm $0.11 & 597 & 2.82 &  4.763$^{e}$ &   8.38 &  2.433 & SH518                & 0.0822 & 11    &   1 \\       
 RXCJ0035.6+0138 & 00 35 36.3 & +01 38 20  & 0.105$\pm $0.022 & 0.75$\pm $0.23 & 291 & 2.66 &  5.000$^{e}$ &   2.13 &        &                      &        &       &     \\       
 RXCJ0037.4-2831 & 00 37 27.0 & -28 31 52  & 0.113$\pm $0.023 & 0.39$\pm $0.21 & 326 & 1.73 &  3.387$^{e}$ &   2.22 &  1.057 & A2798                & 0.1050 &  3    &  20 \\       
 RXCJ0041.8-0918 & 00 41 50.0 & -09 18 12  & 3.554$\pm $0.114 & 0.63$\pm $0.03 & 328 & 3.58 &  6.200$^{ }$ &  74.00 &  9.810 & A85                  & 0.0556 & 12    & 116 \\       
 RXCJ0042.1-2832 & 00 42 08.8 & -28 32 12  & 0.498$\pm $0.049 & 0.28$\pm $0.07 & 325 & 1.49 &  6.768$^{e}$ &   9.82 &  5.726 & A2811                & 0.1170 &  8    &  32 \\       
 RXCJ0043.5-0443 & 00 43 34.6 & -04 43 06  & 0.119$\pm $0.018 & 0.64$\pm $0.13 & 446 & 3.63 &  5.000$^{e}$ &   2.48 &        &                      &        &       &     \\       
 RXCJ0048.6-2114 & 00 48 38.0 & -21 14 50  & 0.217$\pm $0.032 & 0.09$\pm $0.13 & 339 & 1.52 &  2.759$^{e}$ &   4.20 &  0.614 & A2824                & 0.0581 & 17    &   7 \\       
 RXCJ0049.3-2931 & 00 49 23.8 & -29 31 27  & 0.265$\pm $0.042 & 0.40$\pm $0.13 & 323 & 1.80 &  5.172$^{e}$ &   5.25 &  2.974 & S84                  & 0.1150 &  8    &  20 \\       
 RXCJ0051.1-4833 & 00 51 11.2 & -48 33 35  & 0.123$\pm $0.027 & 0.63$\pm $0.24 & 204 & 2.12 &  5.646$^{e}$ &   2.46 &  3.684 & A2830                & 0.1873 &  1    &   2 \\       
 RXCJ0051.3-2831 & 00 51 19.2 & -28 31 10  & 0.127$\pm $0.023 & 0.40$\pm $0.16 & 324 & 1.63 &  3.750$^{e}$ &   2.49 &  1.359 & A2829                & 0.1125 & 13    &   6 \\       
 RXCJ0056.0-3732 & 00 56 01.0 & -37 32 45  & 0.425$\pm $0.034 & 0.55$\pm $0.21 & 351 & 2.59 &  8.538$^{e}$ &   8.65 & 10.102 &                      & 0.1663 &  2    &  11 \\       
 RXCJ0056.3-0112 & 00 56 18.3 & -01 12 53  & 1.644$\pm $0.087 & 0.74$\pm $0.06 & 321 & 3.10 &  5.100$^{ }$ &  33.70 &  2.855 & A119                 & 0.0444 & 17    &  23 \\       
 RXCJ0058.4-1425 & 00 58 28.2 & -14 25 39  & 0.108$\pm $0.018 & 0.17$\pm $0.15 & 492 & 1.87 &  3.105$^{e}$ &   2.12 &  0.840 & A123                 & 0.0957 &  5    &   2 \\       
 RXCJ0102.7-2152 & 01 02 42.3 & -21 52 46  & 0.970$\pm $0.065 & 0.33$\pm $0.06 & 268 & 1.48 &  3.800$^{ }$ &  18.90 &  2.574 & A133                 & 0.0562 &  8    &   9 \\       
 RXCJ0104.5-2400 & 01 04 35.0 & -24 00 00  & 0.119$\pm $0.021 & 0.64$\pm $0.19 & 335 & 1.61 &  4.672$^{e}$ &   2.34 &  2.322 & A140                 & 0.1520 &  8    &  12 \\       
 RXCJ0105.5-2439 & 01 05 35.4 & -24 39 33  & 0.196$\pm $0.031 & 0.32$\pm $0.12 & 326 & 1.61 &  8.018$^{e}$ &   3.88 &  8.635 & A141                 & 0.2300 &  8    &   1 \\       
 RXCJ0106.8-0229 & 01 06 51.5 & -02 29 18  & 0.152$\pm $0.032 & 0.91$\pm $0.23 & 423 & 4.10 &  6.386$^{e}$ &   3.21 &  4.972 & A145                 & 0.1909 &  5    &   1 \\       
 RXCJ0107.8-3643 & 01 07 49.4 & -36 43 44  & 0.180$\pm $0.024 & 0.82$\pm $0.16 & 378 & 1.94 &  4.642$^{e}$ &   3.57 &  2.285 & A2871                & 0.1221 &  2    &  11 \\       
 RXCJ0108.2+0210 & 01 08 12.8 & +02 10 48  & 0.230$\pm $0.037 & 0.59$\pm $0.15 & 415 & 3.01 &  2.342$^{e}$ &   4.60 &  0.398 & A147                 & 0.0447 &  3    &  11 \\       
 RXCJ0108.5-4021 & 01 08 33.2 & -40 21 00  & 0.150$\pm $0.035 & 0.40$\pm $0.23 & 377 & 3.04 &  4.898$^{e}$ &   3.07 &  2.610 & A2874                & 0.1408 & 13    &   6 \\       
 RXCJ0108.8-1524 & 01 08 49.3 & -15 24 30  & 0.358$\pm $0.035 & 0.29$\pm $0.08 & 446 & 1.69 &  3.149$^{e}$ &   7.00 &  0.872 & A151                 & 0.0537 &  8    &  37 \\       
 RXCJ0108.9-1537 & 01 08 55.5 & -15 37 42  & 0.189$\pm $0.029 & 0.46$\pm $0.09 & 437 & 1.69 &  3.954$^{e}$ &   3.72 &  1.541 &                      & 0.0981 &  4,19 &   9 \\       
 RXCJ0110.0-4555 & 01 10 00.7 & -45 55 24  & 0.756$\pm $0.060 & 0.26$\pm $0.08 & 301 & 2.10 &  3.500$^{ }$ &  15.00 &  0.404 & A2877                & 0.0250 &  8    & 109 \\       
 RXCJ0113.9-3145 & 01 13 54.2 & -31 45 07  & 0.080$\pm $0.017 & 0.58$\pm $0.24 & 333 & 2.40 &  1.000$^{e}$ &   1.21 &  0.018 & S141                 & 0.0184 &  8    &  18 \\       
 RXCJ0114.7-2118 & 01 14 47.3 & -21 18 02  & 0.082$\pm $0.018 & 0.56$\pm $0.24 & 346 & 1.49 &  5.000$^{e}$ &   1.61 &        &                      &        &       &     \\       
 RXCJ0114.9+0024 & 01 14 57.5 & +00 24 20  & 0.520$\pm $0.048 & 0.47$\pm $0.09 & 406 & 3.32 &  2.600$^{ }$ &  10.60 &  1.043 & A168                 & 0.0477 &  1    &   2 \\       
 RXCJ0116.1-1555 & 01 16 11.8 & -15 55 24  & 0.130$\pm $0.023 & 0.23$\pm $0.17 & 394 & 1.69 &  1.862$^{e}$ &   2.46 &  0.214 & SCG 16               & 0.0448 &  1    &   2 \\       
 RXCJ0118.1-2658 & 01 18 10.2 & -26 58 18  & 0.193$\pm $0.027 & 0.54$\pm $0.14 & 366 & 1.47 &  7.987$^{e}$ &   3.81 &  8.554 & A2895                & 0.2310 &  1    &   1 \\       
 RXCJ0120.9-1351 & 01 20 57.9 & -13 51 19  & 0.425$\pm $0.034 & 0.30$\pm $0.07 & 344 & 1.73 &  3.240$^{e}$ &   8.33 &  0.939 & CID 10               & 0.0511 &  1    &   3 \\       
 RXCJ0121.8-2022 & 01 21 48.1 & -20 22 34  & 0.110$\pm $0.020 & 0.70$\pm $0.28 & 465 & 1.45 &  5.000$^{e}$ &   2.16 &        & A2902                &        &       &     \\       
 RXCJ0122.2-2131 & 01 22 12.4 & -21 31 04  & 0.097$\pm $0.020 & 0.05$\pm $0.17 & 494 & 1.39 &  5.000$^{e}$ &   1.90 &        & A185                 &        &       &     \\       
 RXCJ0122.9-1245 & 01 22 55.1 & -12 45 50  & 0.096$\pm $0.018 & 0.87$\pm $0.25 & 391 & 1.68 &  3.598$^{e}$ &   1.88 &  1.229 & A188                 & 0.1230 &  3    &   2 \\       
 RXCJ0125.4+0145 & 01 25 29.9 & +01 45 46  & 0.425$\pm $0.034 & 0.77$\pm $0.15 & 413 & 3.08 &  1.446$^{e}$ &   8.03 &  0.107 & NGC 533              & 0.0176 &  2    &   4 \\       
 RXCJ0125.6-0125 & 01 25 38.0 & -01 25 03  & 0.610$\pm $0.086 & 0.84$\pm $0.14 & 431 & 3.56 &  1.900$^{ }$ &  12.20 &  0.176 & A194                 & 0.0183 &  8    &  81 \\       
 RXCJ0126.0-3758 & 01 26 04.6 & -37 58 36  & 0.105$\pm $0.019 & 0.76$\pm $0.23 & 351 & 1.68 &  2.703$^{e}$ &   2.04 &  0.581 & A2911                & 0.0810 &  4    &  31 \\       
 RXCJ0126.7-1810 & 01 26 44.4 & -18 10 20  & 0.103$\pm $0.025 & 0.32$\pm $0.21 & 328 & 1.62 &  5.000$^{e}$ &   2.03 &        & A197                 &        &       &     \\       
 RXCJ0127.2-1746 & 01 27 13.9 & -17 46 17  & 0.109$\pm $0.021 & 0.28$\pm $0.19 & 351 & 1.52 &  4.358$^{e}$ &   2.14 &  1.960 & A199                 & 0.1459 &  1    &   2 \\       
 RXCJ0131.7+0033 & 01 31 42.4 & +00 33 43  & 0.088$\pm $0.017 & 0.26$\pm $0.18 & 432 & 2.92 &  2.535$^{e}$ &   1.77 &  0.490 & A208                 & 0.0798 &  3    &   ? \\       
 RXCJ0131.8-1336 & 01 31 53.5 & -13 36 27  & 0.277$\pm $0.029 & 0.33$\pm $0.10 & 453 & 2.27 &  8.513$^{e}$ &   5.59 &  9.982 & A209                 & 0.2060 &  3    &   2 \\       
 RXCJ0137.2-0912 & 01 37 15.3 & -09 12 10  & 0.425$\pm $0.034 & 0.68$\pm $0.08 & 444 & 2.75 &  2.672$^{e}$ &   8.50 &  0.564 &                      & 0.0392 &  1    &   3 \\       
 RXCJ0137.4-1259 & 01 37 29.2 & -12 59 10  & 0.088$\pm $0.016 & 0.55$\pm $0.18 & 458 & 2.45 &  5.503$^{e}$ &   1.78 &  3.462 & A222                 & 0.2134 &  8    &   6 \\       
 RXCJ0137.9-1248 & 01 37 56.4 & -12 48 01  & 0.115$\pm $0.018 & 0.60$\pm $0.20 & 454 & 1.84 &  5.935$^{e}$ &   2.28 &  4.160 & A223                 & 0.2070 &  8    &   4 \\       
 RXCJ0139.1-1915 & 01 39 10.1 & -19 15 34  & 0.083$\pm $0.015 & 0.28$\pm $0.20 & 479 & 1.45 &  5.000$^{e}$ &   1.63 &        &                      &        &       &     \\       
 RXCJ0139.9-0555 & 01 39 59.7 & -05 55 23  & 0.084$\pm $0.018 & 0.43$\pm $0.20 & 441 & 2.85 &  5.000$^{e}$ &   1.71 &        &                      &        &       &     \\       
 RXCJ0143.4-4614 & 01 43 29.8 & -46 14 20  & 0.082$\pm $0.017 & 0.41$\pm $0.25 & 455 & 2.30 &  5.000$^{e}$ &   1.65 &        & A2937                &        &       &     \\       
 RXCJ0144.6-2213 & 01 44 41.8 & -22 13 43  & 0.097$\pm $0.015 & 0.25$\pm $0.13 & 494 & 1.18 &  6.991$^{e}$ &   1.90 &  6.196 & A2938                & 0.2781 &  1    &   1 \\       
 RXCJ0148.2-3155 & 01 48 14.8 & -31 55 08  & 0.105$\pm $0.016 & 0.37$\pm $0.17 & 521 & 1.53 &  3.400$^{ }$ &   2.05 &        & A2943                &        &       &     \\       
 RXCJ0148.3-0406 & 01 48 22.2 & -04 06 13  & 0.104$\pm $0.024 & 0.25$\pm $0.19 & 255 & 2.59 &  5.000$^{e}$ &   2.10 &        &                      &        &       &     \\       
 RXCJ0152.5-2853 & 01 52 32.8 & -28 53 32  & 0.084$\pm $0.016 & 0.16$\pm $0.16 & 391 & 1.51 &  5.000$^{e}$ &   1.65 &        &                      &        &       &     \\       
 RXCJ0152.7+0100 & 01 52 43.6 & +01 00 58  & 0.151$\pm $0.022 & 0.42$\pm $0.13 & 407 & 2.84 &  7.309$^{e}$ &   3.09 &  6.900 & A267 dc              & 0.2300 &  9    &   1 \\       
 RXCJ0152.9-1345 & 01 52 59.8 & -13 45 07  & 0.425$\pm $0.034 & 0.25$\pm $0.13 & 462 & 1.69 &  1.000$^{e}$ &   6.30 &  0.009 & NGC 720              & 0.0057 & 15    &   1 \\       
 RXCJ0153.5-0118 & 01 53 32.0 & -01 18 44  & 0.123$\pm $0.030 & 0.54$\pm $0.17 & 433 & 2.66 &  7.025$^{e}$ &   2.50 &  6.277 &                      & 0.2438 &  1    &   2 \\       
 RXCJ0157.5-0549 & 01 57 30.5 & -05 49 32  & 0.164$\pm $0.024 & 0.55$\pm $0.19 & 414 & 2.31 &  4.681$^{e}$ &   3.29 &  2.332 & A281 dc              & 0.1285 &  2    &   2 \\       
 RXCJ0158.4-0146 & 01 58 28.4 & -01 46 51  & 0.116$\pm $0.020 & 0.48$\pm $0.17 & 406 & 2.55 &  4.948$^{e}$ &   2.34 &  2.672 & A286                 & 0.1632 &  1    &   1 \\       
 RXCJ0201.7-0211 & 02 01 44.3 & -02 11 58  & 0.209$\pm $0.024 & 0.52$\pm $0.11 & 406 & 2.57 &  7.327$^{e}$ &   4.25 &  6.941 & A291                 & 0.1965 &  9    &   1 \\       
 RXCJ0202.3-4447 & 02 02 18.8 & -44 47 37  & 0.124$\pm $0.020 & 0.31$\pm $0.18 & 465 & 2.70 &  2.359$^{e}$ &   2.46 &  0.405 &                      & 0.0616 &  1    &   2 \\       
 RXCJ0202.3-0107 & 02 02 20.6 & -01 07 06  & 0.191$\pm $0.026 & 0.22$\pm $0.13 & 405 & 2.58 &  2.103$^{e}$ &   3.76 &  0.298 & A295                 & 0.0428 &  5    &   3 \\       
 RXCJ0206.4-1453 & 02 06 29.9 & -14 53 37  & 0.149$\pm $0.027 & 0.52$\pm $0.16 & 274 & 2.45 &  5.192$^{e}$ &   3.00 &  3.003 & A305                 & 0.1529 &  1    &   2 \\       
 RXCJ0208.0-1537 & 02 08 00.2 & -15 37 05  & 0.085$\pm $0.019 & 0.92$\pm $0.24 & 286 & 2.38 &  5.000$^{e}$ &   1.71 &        &                      &        &       &     \\       
 RXCJ0211.4-4017 & 02 11 25.2 & -40 17 09  & 0.166$\pm $0.019 & 0.54$\pm $0.14 & 583 & 1.43 &  3.859$^{e}$ &   3.24 &  1.457 & A2984                & 0.1022 &  8    &   6 \\       
 RXCJ0212.8-4707 & 02 12 53.9 & -47 07 58  & 0.197$\pm $0.026 & 0.35$\pm $0.18 & 517 & 2.12 &  4.598$^{e}$ &   3.93 &  2.232 & A2988                & 0.1150 & 25    &   ? \\       
 RXCJ0213.9-0253 & 02 13 56.6 & -02 53 53  & 0.101$\pm $0.022 & 0.46$\pm $0.20 & 350 & 2.16 &  5.000$^{e}$ &   2.02 &        &                      &        &       &     \\       
 RXCJ0214.6-0433 & 02 14 41.1 & -04 33 48  & 0.106$\pm $0.021 & 0.68$\pm $0.30 & 357 & 2.21 &  4.181$^{e}$ &   2.12 &  1.772 & A329                 & 0.1393 &  1    &   2 \\       
 RXCJ0216.3-4816 & 02 16 19.1 & -48 16 23  & 0.140$\pm $0.019 & 0.85$\pm $0.23 & 542 & 2.95 &  5.573$^{e}$ &   2.86 &  3.568 & A2998                & 0.1709 &  1    &   1 \\       
 RXCJ0216.7-4749 & 02 16 43.3 & -47 49 31  & 0.180$\pm $0.030 & 0.89$\pm $0.28 & 523 & 2.95 &  2.809$^{e}$ &   3.63 &  0.644 & S239                 & 0.0640 & 23    &   1 \\       
 RXCJ0218.3-3141 & 02 18 19.6 & -31 41 34  & 0.099$\pm $0.016 & 0.28$\pm $0.14 & 486 & 1.87 &  5.000$^{e}$ &   1.96 &        &                      &        &       &     \\       
 RXCJ0219.1-1724 & 02 19 08.3 & -17 24 40  & 0.085$\pm $0.016 & 0.82$\pm $0.23 & 390 & 2.99 &  5.000$^{e}$ &   1.74 &        &                      &        &       &     \\       
 RXCJ0220.9-3829 & 02 20 56.8 & -38 29 02  & 0.425$\pm $0.034 & 0.74$\pm $0.17 & 558 & 1.85 & 10.965$^{e}$ &   8.48 & 18.532 &                      & 0.2287 &  2    &   6 \\       
 RXCJ0225.1-2928 & 02 25 10.2 & -29 28 23  & 0.425$\pm $0.034 & 0.63$\pm $0.15 & 445 & 1.70 &  3.710$^{e}$ &   8.35 &  1.327 &                      & 0.0607 &  1    &   2 \\       
 RXCJ0225.8-4154 & 02 25 53.7 & -41 54 19  & 0.258$\pm $0.029 & 0.29$\pm $0.10 & 460 & 2.14 &  8.693$^{e}$ &   5.19 & 10.503 & A3017                & 0.2195 &  1    &   1 \\       
 RXCJ0227.2-2851 & 02 27 12.6 & -28 51 18  & 0.071$\pm $0.018 & 0.59$\pm $0.21 & 343 & 1.55 &  5.003$^{e}$ &   1.40 &  2.743 & EDCC651              & 0.2138 &  2    &   ? \\       
 RXCJ0229.3-3332 & 02 29 22.3 & -33 32 13  & 0.198$\pm $0.026 & 0.34$\pm $0.12 & 471 & 2.07 &  3.352$^{e}$ &   3.92 &  1.029 & APMCC269             & 0.0779 &  1    &   2 \\       
 RXCJ0229.9-1316 & 02 29 56.4 & -13 16 12  & 0.127$\pm $0.026 & 0.25$\pm $0.20 & 312 & 2.04 &  5.000$^{e}$ &   2.53 &        &                      &        &       &     \\       
 RXCJ0230.8-3305 & 02 30 51.2 & -33 05 49  & 0.166$\pm $0.026 & 0.71$\pm $0.24 & 456 & 1.96 &  3.116$^{e}$ &   3.27 &  0.848 & A3027                & 0.0774 & 13    &   4 \\       
 RXCJ0231.7-0451 & 02 31 47.3 & -04 51 21  & 0.124$\pm $0.033 & 0.55$\pm $0.24 & 125 & 2.55 &  5.619$^{e}$ &   2.51 &  3.640 & A362                 & 0.1843 &  1    &   3 \\       
 RXCJ0231.9+0114 & 02 31 57.5 & +01 14 38  & 0.425$\pm $0.034 & 0.97$\pm $0.26 & 227 & 2.56 &  1.714$^{e}$ &   8.12 &  0.171 &                      & 0.0221 &  1    &   2 \\       
 RXCJ0232.2-4420 & 02 32 17.2 & -44 20 43  & 0.425$\pm $0.034 & 0.14$\pm $0.09 & 369 & 2.61 & 13.183$^{e}$ &   8.66 & 28.933 &                      & 0.2836 &  1    &   2 \\       
 RXCJ0236.6-1923 & 02 36 39.6 & -19 23 05  & 0.171$\pm $0.025 & 0.57$\pm $0.19 & 349 & 2.65 &  4.100$^{ }$ &   3.45 &  1.179 & A367                 & 0.0891 & 17    &   3 \\       
 RXCJ0237.4-2630 & 02 37 29.0 & -26 30 13  & 0.156$\pm $0.024 & 0.43$\pm $0.16 & 325 & 1.56 &  7.042$^{e}$ &   3.08 &  6.305 & A368                 & 0.2200 &  2    &   1 \\       
 RXCJ0241.3-2839 & 02 41 20.9 & -28 39 27  & 0.130$\pm $0.021 & 0.10$\pm $0.13 & 423 & 1.56 &  4.300$^{ }$ &   2.55 &  5.942 & A3041                & 0.2323 &  1    &   3 \\       
 RXCJ0244.1-2611 & 02 44 06.3 & -26 11 07  & 0.152$\pm $0.027 & 0.79$\pm $0.25 & 286 & 1.74 &  4.726$^{e}$ &   3.00 &  2.389 &                      & 0.1362 &  4,19 &   6 \\       
 RXCJ0245.2-4627 & 02 45 12.6 & -46 27 14  & 0.114$\pm $0.017 & 0.66$\pm $0.22 & 539 & 3.15 &  2.986$^{e}$ &   2.32 &  0.757 & A3047                & 0.0868 &  1    &   1 \\       
 RXCJ0248.0-0332 & 02 48 02.6 & -03 32 09  & 0.229$\pm $0.042 & 0.49$\pm $0.16 & 224 & 4.12 &  7.521$^{e}$ &   4.85 &  7.396 & A383                 & 0.1899 &  1    &   1 \\       
 RXCJ0248.2-0216 & 02 48 16.3 & -02 16 44  & 0.220$\pm $0.033 & 0.66$\pm $0.22 & 117 & 3.65 &  8.748$^{e}$ &   4.60 & 10.743 & A384                 & 0.2360 &  2    &   ? \\       
 RXCJ0249.6-3111 & 02 49 36.9 & -31 11 19  & 0.469$\pm $0.040 & 1.31$\pm $0.27 & 419 & 1.80 &  1.837$^{e}$ &   8.89 &  0.206 & S301                 & 0.0232 &  8    &  34 \\       
 RXCJ0251.4-2456 & 02 51 28.8 & -24 56 58  & 0.144$\pm $0.023 & 0.04$\pm $0.12 & 365 & 1.90 &  4.500$^{ }$ &   2.85 &  1.511 & A389                 & 0.1111 &  8    &   6 \\       
 RXCJ0251.7-4109 & 02 51 42.3 & -41 09 00  & 0.102$\pm $0.018 & 0.00$\pm $0.17 & 496 & 1.86 &  5.000$^{e}$ &   2.02 &        &                      &        &       &     \\       
 RXCJ0252.8-0116 & 02 52 50.1 & -01 16 28  & 0.425$\pm $0.034 & 1.34$\pm $0.23 & 135 & 5.25 &  1.833$^{e}$ &   8.84 &  0.205 & NGC 1132             & 0.0232 & 10    &   1 \\       
 RXCJ0258.2-2105 & 02 58 15.0 & -21 05 29  & 0.114$\pm $0.023 & 0.57$\pm $0.21 & 287 & 1.92 &  3.859$^{e}$ &   2.25 &  1.458 & A3073                & 0.1226 &  1    &   1 \\       
 RXCJ0303.2-2735 & 03 03 14.8 & -27 35 59  & 0.122$\pm $0.028 & 0.25$\pm $0.20 & 266 & 1.55 &  5.000$^{e}$ &   2.40 &        & A3082                &        &       &     \\       
 RXCJ0303.3+0155 & 03 03 21.3 & +01 55 35  & 0.219$\pm $0.033 & 0.83$\pm $0.15 & 222 & 7.82 &  6.421$^{e}$ &   5.08 &  5.062 & A409                 & 0.1530 &  9    &   1 \\       
 RXCJ0304.0-3656 & 03 04 06.0 & -36 56 32  & 0.151$\pm $0.023 & 0.76$\pm $0.23 & 384 & 1.95 &  6.958$^{e}$ &   3.01 &  6.120 & A3084                & 0.2192 &  2    &   ? \\       
 RXCJ0307.0-2840 & 03 07 04.0 & -28 40 15  & 0.197$\pm $0.034 & 0.38$\pm $0.15 & 237 & 1.36 &  8.674$^{e}$ &   3.88 & 10.434 & A3088                & 0.2534 &  7    &   ? \\       
 RXCJ0311.4-2654 & 03 11 25.3 & -26 54 00  & 0.154$\pm $0.078 & 0.52$\pm $0.22 & 210 & 1.56 &  2.722$^{e}$ &   2.98 &  0.593 & A3094/EDCC736 dc     & 0.0677 &  3    &  67 \\       
 RXCJ0313.6-3817 & 03 13 38.0 & -38 17 55  & 0.119$\pm $0.023 & 0.20$\pm $0.22 & 321 & 2.03 &  2.905$^{e}$ &   2.34 &  0.704 & A3098                & 0.0833 &  3    &   6 \\       
 RXCJ0314.3-4525 & 03 14 19.6 & -45 25 24  & 0.401$\pm $0.038 & 0.45$\pm $0.10 & 313 & 3.57 &  4.252$^{e}$ &   8.31 &  1.845 & A3104                & 0.0718 &  2    &   4 \\       
 RXCJ0317.7-4848 & 03 17 46.4 & -48 48 42  & 0.097$\pm $0.016 & 0.56$\pm $0.23 & 487 & 2.06 &  5.000$^{e}$ &   1.93 &        & A3113                &        &       &     \\       
 RXCJ0317.9-4414 & 03 17 58.2 & -44 14 16  & 1.447$\pm $0.056 & 0.30$\pm $0.04 & 481 & 2.53 &  4.100$^{ }$ &  29.10 &  7.130 & A3112                & 0.0754 &  1    &   2 \\ \hline
\enddata                                                                
\end{deluxetable}                                                       
\end{document}